\documentclass[aps,twocolumn,amsmath,amssymb,prx,showpacs]{revtex4-1}

\usepackage{graphicx}
\usepackage{comment}
\usepackage[normalem]{ulem}
\usepackage{color}

\definecolor{darkpastelgreen}{rgb}{0.01, 0.75, 0.24}

\makeatletter
\def\blfootnote{\xdef\@thefnmark{}\@footnotetext}
\makeatother

\begin{document}
%
%
\title{Hall anomaly, Quantum Oscillations and Possible Lifshitz Transitions in Kondo Insulator YbB$_{12}$: Evidence for Unconventional Charge Transport}
\author{Ziji Xiang$^{1,2,*}$, Kuan-Wen Chen$^1$, Lu Chen$^{1,\dag}$, Tomoya Asaba$^{1,3}$, Yuki Sato$^3$, Nan Zhang$^2$, Dechen Zhang$^1$, Yuichi Kasahara$^3$, Fumitoshi Iga$^4$, William A. Coniglio$^5$, Yuji Matsuda$^3$, John Singleton$^{6,*}$, Lu Li$^{1,*}$}
\affiliation{
$^1$Department of Physics, University of Michigan, Ann Arbor, Michigan 48109, USA\\
$^2$CAS Key Laboratory of Strongly-coupled Quantum Matter Physics, Department of Physics, University of Science and Technology of China, Hefei, Anhui 230026, China\\
$^3$Department of Physics, Kyoto University, Kyoto 606-8502, Japan\\
$^4$Institute of Quantum Beam Science, Graduate School of Science and Engineering, Ibaraki University, Mito 310-8512, Japan\\
$^5$National High Magnetic Field Laboratory, 1800 East Paul Dirac Drive, Tallahassee, Florida 32310-3706, USA\\
$^6$National High Magnetic Field Laboratory, MS E536, Los Alamos National Laboratory, Los Alamos, New Mexico 87545, USA
}

\date{\today}
\begin{abstract}
In correlated electronic systems, strong interactions and the interplay between different degrees of freedom may give rise to anomalous charge transport properties, which can be tuned by external parameters like temperature and magnetic field. Recently, magnetic quantum oscillations and metallic low-temperature thermal conductivity have been observed in the Kondo insulator YbB$_{12}$, whose resistivity is a few orders of magnitude higher than those of conventional metals. As yet, these unusual observations are not fully understood. Here we present a detailed investigation of the behavior of YbB$_{12}$ under intense magnetic fields using both transport and torque magnetometry measurements. The Hall resistivity displays a strongly nonlinear field dependence which cannot be described using a standard two-band Drude model.
A low-field Hall anomaly, reminiscent of the Hall response associated with ``strange-metal'' physics, develops at $T < 1.5$~K. At two characteristic magnetic fields
($\mu_0H_1= 19.6$~T and $\mu_0H_2 \sim 31$~T), signatures appear in the Hall coefficient, magnetic torque, and magnetoresistance; the latter characteristic field coincides with the occurrence of a change in quantum oscillation frequency. We suggest that they are likely to be field-induced Lifshitz transitions. Moreover, above 35~T, where the most pronounced quantum oscillations are detected, the background resistivity displays an unusual, nonmetallic $T^{\alpha}$-behavior, with $\alpha$ being field-dependent and varying between -1.5 and -2. By normalizing the Shubnikov-de Haas oscillation amplitude to this $T^{\alpha}$-dependence, the calculated cyclotron mass becomes more consistent with that deduced from de Haas-van Alphen oscillations. Our results support a novel two-fluid scenario in YbB$_{12}$: a Fermi-liquid-like fluid of charge-neutral quasiparticles coexists with charge carriers that remain in a nonmetallic state. The former experience successive Lifshitz transitions and develop Landau quantization in applied magnetic fields, whilst scattering between both fluids allows the Shubnikov-de Haas effect to be observed in the electrical transport. The verification of this two-fluid scenario by the data in the current work strongly suggests that YbB$_{12}$ represents a new phase of matter.
\end{abstract}

\blfootnote{$^*$Corresponding authors: zijixiang@ustc.edu.cn, jsingle@lanl.gov, luli@umich.edu}
\blfootnote{$^\dag$Present address: Institut Quantique, D\'{e}partement de physique and RQMP, Universit\'{e} de Sherbrooke, Sherbrooke, Qu\'{e}bec J1K 2R1, Canada}


\maketitle                   
\renewcommand{\thesubsection}{\Alph{subsection}}
\renewcommand{\thesubsubsection}{\textit{\alph{subsubsection}}}
\section{INTRODUCTION}
\label{chap01}

Kondo lattices are systems containing both conduction electrons (usually itinerant $d$-electrons) and periodic arrays of local magnetic moments (typically localized $f$-electrons). At low temperatures ($T$), interactions between these $d$- and $f$-bands lead to screening of the local moments, forming many-body states called Kondo resonances that in turn hybridize with the conduction-electron states~\cite{HewsonBook}. This $f-d$ hybridization opens a gap between the two hybridized quasiparticle bands that is centered on the energy level of the renormalized $f$-states. If the conduction band is at exact half filling~\cite{Riseborough,ColemanBookChapter}, or, more commonly, if the $f$-states have a mixed valence~\cite{VarmaMixed}, the Fermi energy lies within the hybridization gap, making the system a {\it Kondo insulator} (KI). Archetypal examples of KIs are cubic-structured $f$-electron compounds~\cite{MenthSmB6,KasayaMixed,Hundley433} SmB$_6$, YbB$_{12}$ and Ce$_3$Bi$_4$Pt$_3$. Thanks to the presence of strong electron correlations, the physical properties of these KIs are very unusual, differing significantly from those of conventional band insulators. For instance, the Kondo gap is usually quite narrow (most commonly 5-20 meV) and $T-$dependent~\cite{MenthSmB6,KasayaMixed,Hundley433}, exhibiting in-gap excitations even in the ground state \cite{CaldwellNMR,OkawaTRPES}. In particular, the resistivity of KIs deviates from thermally activated behavior at temperatures of a few Kelvin, below which the resistivity tends to saturate~\cite{KasayaMixed,CooleySmB6}. It is difficult to attribute these low-$T$ resistive ``plateaus" to conventional extended in-gap states because the large residual resistivity suggests unphysically short quasiparticle relaxation times~\cite{CooleySmB6}. Recently, theoretical works predicted
time-reversal-invariant topological-insulator phases in some KIs with simple cubic structure~\cite{dzero2010topological,dzero2012theory,alexandrov2013cubic}. Within such models, the low-$T$ resistive plateau is taken as evidence for two-dimensional (2D) metallic transport channels associated with topological surface states.
The proposed topological Kondo insulators (TKIs) attracted much theoretical and experimental interest, as they provide a new route towards the study of strongly correlated topological systems.

The most intensively studied KI is the mixed-valence compound SmB$_6$, which has a simple cubic lattice and is a candidate TKI~\cite{takimoto2011smb6,lu2013correlated}. Numerous angle-resolved photoemission spectroscopy (ARPES)~\cite{NeupaneARPES2013,JiangARPES2013,xu2014direct,OhtsuboARPES111,RyuARPES110,MiaoARPES}, scanning tunnelling microscopy~\cite{JiaoSTM2016,JiaoSTM2018,PirieSTM}, planar tunneling spectroscopy~\cite{WanKyuPNAS,WanKyuPRB} and electrical transport~\cite{WolgastPRB,kim2014topological} experiments yield data supporting topological surface states in SmB$_6$. However, other experiments can be interpreted as not favouring the TKI scenario~\cite{HlawenkaARPES,Shahrokhvand}. The proposed topological nontriviality of other KIs is as yet inconclusive.

YbB$_{12}$ crystallizes in a face-centered cubic structure and also exhibits mixed valence properties \cite{KasayaMixed}; it is predicted to be a topological crystalline insulator that hosts surface states protected by the lattice symmetry~\cite{YbB12TCI}. However, while magnetotransport measurements confirm the contribution of surface transport channels~\cite{FIBYbB12Sato}, ARPES results suggest that the topological surface states are more consistent with a strong topological insulator, rather than a topological crystalline insulator~\cite{ARPESYbB12}.

More intriguingly, unconventional magnetic quantum oscillations~\cite{tan2015unconventional,Xiang2018SdH,Xiang2021pulsed,HsuLiu} and finite residual thermal conductivity $\kappa_{xx}/T ~~(T\rightarrow0)$~\cite{Hartstein2018Fermi,Sato2019thermal} attributable to three-dimensional (3D) fermionic states have been observed in both SmB$_6$ and YbB$_{12}$. Such observations are surprising since the two compounds exhibit $T\rightarrow 0$ resistivities  that are at least 4 orders of magnitude higher than those of conventional metals; hence, extended electronic states (and therefore a Fermi surface) are not expected. The case of SmB$_6$ is controversial, however, because the quantum oscillations~\cite{li2014two,Xiang2017,ThomasAlFlux} and the presence/absence of residual thermal conductivity \cite{SYLi,Boulanger} differ in single crystals grown by different techniques. Furthermore, only the de Haas-van Alphen (dHvA) effect ({\it i.e.,} quantum oscillations in magnetization) has been reported in SmB$_6$, while the Shubnikov-de Haas (SdH) effect ({\it i.e.,} quantum oscillations in electrical resistivity) has not been observed. All of these inconsistencies lead to doubts that the unconventional properties of SmB$_6$ are intrinsic~\cite{Li2021review}.

Both SdH and dHvA oscillations are observed in YbB$_{12}$; these oscillations are well described by the Lifshitz-Kosevich (LK) formula that is appropriate for Fermi liquids. Moreover, the various extrinsic explanations for the observed phenomena can be excluded~\cite{Xiang2018SdH,Xiang2021pulsed}. The Fermi-surface (FS) sections revealed by the quantum oscillations are 3D pockets containing quasiparticles with relatively large effective masses~\cite{Xiang2018SdH}; these are qualitatively consistent with the itinerant fermion contribution to the low$-T$ thermal conductivity~\cite{Sato2019thermal}. These results provoke a natural interpretation that the ``insulating" bulk of YbB$_{12}$ hosts exotic quasiparticles: they couple to magnetic fields and are able to carry thermal energy, but have a very weak response to electric fields. Recently, we discovered that the SdH effect of these unconventional fermions survives {\it above} the insulator-metal (I-M) transition at a magnetic field $(H)$ higher than $\mu_0H \approx 45~$T, yet changes its behavior in the high-field Kondo-metal (KM) state~\cite{Xiang2021pulsed}. The complex behavior of YbB$_{12}$ strongly suggests that the Landau quantization occurs in a band of charge-neutral quasiparticles which also give rise to the metallic $\kappa_{xx}$ \cite{Xiang2021pulsed}.

The presence of charge-neutral fermions in YbB$_{12}$ suggests that, in addition to the postulated topological surface states, the bulk states of KIs are also unusual and may provide another approach to strongly correlated topological quantum matter. In a mean-field description of the Kondo lattice, the local moments can be fractionalized into multiplets of Majorana fermions, resulting in gapless 3D Majorana channels even in a charge-insulating ground state~\cite{Coleman1993,Coleman1994Odd,Coleman1995Simp,erten2017skyrme}. More recent theoretical works consider topologically protected gapless Majoranas that stem from the phase shifts of bands in mixed valence insulators~\cite{VarmaMajorana} or a Landau-Fermi liquid ground state containing Majorana polarons in KIs~\cite{MajoranaPolaron}. In addition to the latter proposals, other models propose spin-charge separation in mixed valence materials~\cite{Sodemann,Chowdhury,RaoCyclotron}: emergent neutral fermions that couple to a U(1) gauge field (similar to the spinons in quantum-spin liquids~ \cite{Motrunich}) can appear in such a scenario and subsequently hybridize into fermionic composite excitons~\cite{Chowdhury}. Given these competing theories of charge-neutral fermions, further experimental data are required to clarify the nature of the ground state in YbB$_{12}$. In particular, a key question to be addressed is that how these change-neutral fermions lead to the quantum oscillations in electrical transport properties.

In this work, we report a detailed study of the Hall effect, magnetoresistance (MR) and magnetic torque ($\tau$) in high-quality single crystals of YbB$_{12}$ in static magnetic fields of up to 45\,T and pulsed magnetic fields of up to 55\,T. Our main findings include the following.
\begin{enumerate}
\item
A low-field anomaly in the Hall resistivity $\rho_{xy}$ appears at $T \lesssim1.5~$K and becomes almost $T$-independent for $T <0.9~$K. This anomaly cannot be described by a two-band Drude model that takes into account surface conduction. However, it resembles the Hall response in strange metals without full quasiparticle coherence~\cite{PutzkeHall,HallBaFeAsP,HallFeSeS}.
\item
Successive characteristic magnetic fields are indicated by features in $\rho_{xy}$, MR and $\tau$ at $\mu_0H$ = 19.6\,T and $\simeq$ 31\,T. We suggest these are signatures of putative field-induced Lifshitz transitions in the FS sections of the unconventional neutral quasiparticles.
\item
The SdH effect is observed in the field range ${\rm 25.5~T \leq \mu_0H \leq 30~T}$ for the first time. The oscillations possess smaller frequencies but suggest cyclotron masses comparable to those associated with the SdH effect measured above 35\,T~\cite{Xiang2018SdH,Sato2019thermal}. This further supports the occurrence of electronic transition(s).
\item
The previously reported difference between the cyclotron masses extracted
from dHvA and SdH data~\cite{Xiang2018SdH} can be partly resolved by considering the unusual $T^\alpha$-dependence ($\alpha <$ 0) of the background resistivity in high magnetic fields. This $T$ dependence suggests one or more novel scattering mechanisms in this system.
\end{enumerate}
These results strongly imply that any ``conventional'' electrons and holes ({\it i.e.,} charge carriers) in the Kondo-insulator state of YbB$_{12}$, either on the surface or in the bulk, form a possibly non-Fermi-liquid (non-FL) state that dominates the unusual transport behavior. By contrast, there is also a FL-like state of neutral qusiparticles that experience orbital quantization under magnetic fields. Consequently, the SdH effect appears through scattering between the neutral quasiparticles and the charged excitations. Such a scenario points to a unique two-fluid picture in YbB$_{12}$ incorporating both emergent exotic excitations and potential strange metal physics, as well as the interplay between them. YbB$_{12}$ is thus an ideal platform for investigating unusual many-body effects in condensed-matter physics.

\begin{figure}[htbp!]
\centering
\includegraphics[width=0.46\textwidth]{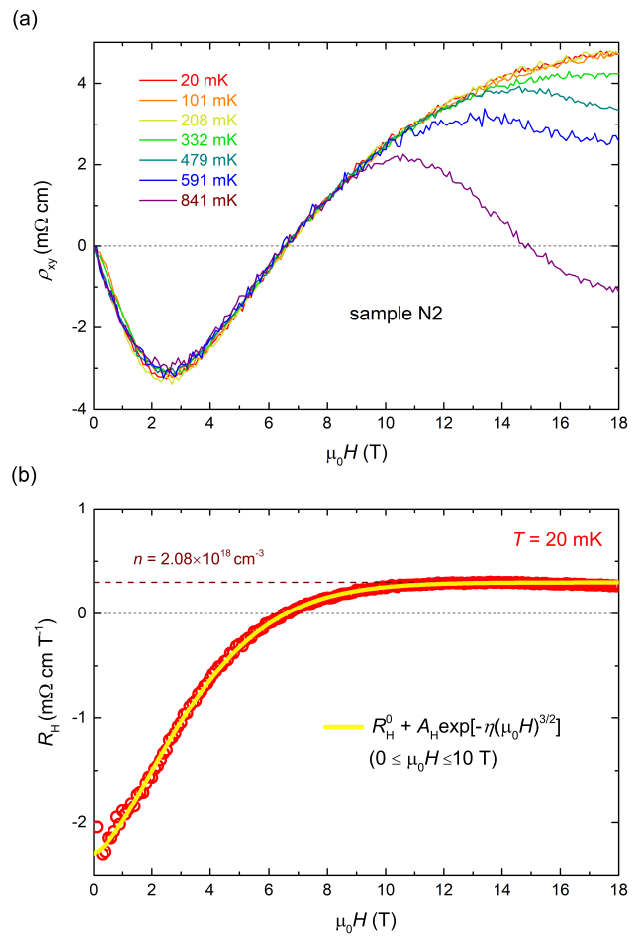}
\caption{(a) Hall resistivity $\rho_{xy}$ of YbB$_{12}$ sample N2 measured in a dilution refrigerator in magnetic fields of up to 18 T. $\rho_{xy}(H)$ shows a sign change at $\mu_0H \simeq 6.6~$T. Below this field, $\rho_{xy}$ is negative and highly nonlinear with magnetic field, showing a low-field Hall anomaly which is nearly $T$-independent from 20~mK to 841~mK. (b)~Fitting of the field dependence of the Hall coefficient $R_{\rm H}(H) \equiv \rho_{xy}(H)/(\mu_0 H)$ below 10 T using (yellow thick line) the empirical expression Eq.\,\ref{empirical}.
}
\label{FigHallSCM1}
\end{figure}

\section{METHODS}
\label{chap02}

The single-crystal YbB$_{12}$ samples were grown at the Ibaraki University using the floating zone method~\cite{Iga}. Samples were cut from the ingots and polished into rectangular bar-shape platelets with all surfaces normal to the $[100]$ principal axes of the cubic lattice. Two batches of samples were measured here. The first batch includes samples N1 and N2, which were characterized in our previous works using quantum-oscillation and thermal-transport measurements~\cite{Xiang2018SdH,Sato2019thermal,Xiang2021pulsed}. The typical dimensions of these samples are $2.3 \times 1 \times 0.2~$mm$^3$ (length $\times$ width $\times$ thickness). The second batch, including samples Ha and Hb, has typical dimensions $2.5 \times 0.6 \times 0.1~$mm$^3$. No differences in physical properties were observed within each batch of samples. However, the behavior of the low-$T$ resistivity differs slightly between the two batches (see Appendix Sec.\,\ref{chap07:1}).

The longitudinal resistivity $\rho_{xx}$ and the Hall resistivity $\rho_{xy}$ were measured simultaneously in a conventional six-probe Hall-bar configuration. Gold wires with diameter 25 $\mu$m were attached to the sample surface using Dupont 4922N silver paint, resulting in contact resistances $\sim 1 \Omega$.. In pulsed-field Hall measurements, the contacts of the voltage leads were further secured by a small amount of Epo-Tek H20E silver epoxy for stability against mechanical vibrations during pulses. Hall resistivity data were taken in three different cryostats: a dilution refrigerator in an 18 T superconducting magnet (SCM1) at the National High Magnetic Field Laboratory (NHMFL) in Tallahassee (sample N2), a Quantum Design DynaCool-14 T system with $^3$He insert (sample N1) and a $^3$He cryostat in a capacitor-driven 65\,T pulsed magnet at the NHMFL in Los Alamos (samples Ha and Hb). To remove stray $\rho_{xx}$ contributions from the Hall voltage, $\rho_{xy}$ was extracted using $\rho_{xy} = t\cdot [V_y(+H) - V_y(-H)]/2I_x$, where $(-H)$ and $(+H)$ indicate values obtained on negative and positive field sweeps and $V_y$ and $I$ are the measured voltage and applied electric current, respectively; $t$ is the sample thickness. In the pulsed-field Hall experiments, this procedure was done separately for up- and down-sweeps of the field to eliminate possible effects of inductive (eddy-current) heating.

Torque magnetometry measurements were performed in magnetic fields of up to 45~T in the hybrid magnet at NHMFL, Tallahassee. The details of the procedures used are given in one of our earlier works~\cite{Xiang2018SdH}.

\section{The Low-field Hall Anomaly}
\label{chap03}
\subsection{The low-field Hall effect in YbB$_{12}$}
\label{chap03:1}

Figure 1(a) shows $\rho_{xy}(H)$ for YbB$_{12}$ sample N2 at $T$s ranging from 20~mK to 841~mK measured at fields of up to $\mu_0H$ = 18\,T. The field-dependence of $\rho_{xy}$  is quite unlike those of conventional insulators or doped semiconductors. First, at these low $T$s, $\rho_{xy}$($H$) is almost $T$-independent below $9~$T, suggesting that the low-$T$, low-$H$ Hall effect is {\it not} primarily due to thermally-activated charge carriers. Second, $\rho_{xy}$ does not vary linearly with $H$. At low $H$, $\rho_{xy}$ is negative ({\it i.e.,} dominated by negative charge carriers); subsequently, following a minimum at $\approx 2.6~$T, it increases, changing sign at 6.6\,T. Furthermore, at $T$ = 841 mK, $\rho_{xy}$($H$) again changes sign (from positive to negative) at $\approx 14.8~$T. This behaviour contrasts sharply with that in SmB$_6$ (where $\rho_{xy} \propto H$~\cite{YLuoSmB6}), and suggests a complicated multiband character for the low-field charge transport in YbB$_{12}$. Conventionally, such a field-dependent Hall effect would be understood in terms of a metal/semimetal with multiple electron and hole FS pockets. However, in YbB$_{12}$ the large, nonmetallic low-$T$ $\rho_{xx}$ (Appendix Sec.\,\ref{chap07:1}) and the strongly negative MR (Appendix Sec.\,\ref{chap07:2}) preclude the existence of a bulk FS comprising charged quasiparticles. We discuss the validity of applying a conventional band-transport picture to YbB$_{12}$ Hall data in more detail in the Appendix (Sec.\,\ref{chap07:3}), where we show that the nonlinear $\rho_{xy}$ displayed in Fig.\,1(a) cannot be interpreted within the framework of a conventional two-band Drude model.

\begin{figure}[htbp!]
\centering
\includegraphics[width=0.46\textwidth]{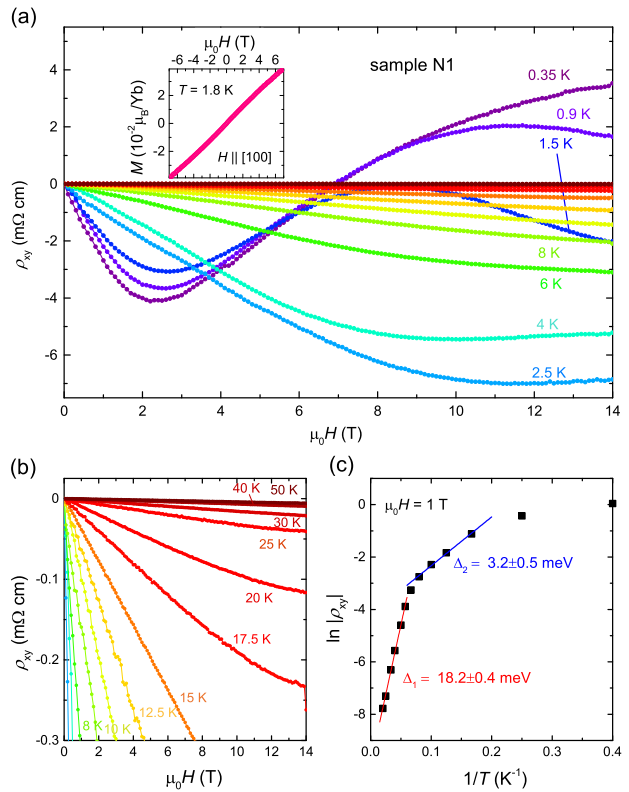}
\caption{(a) $\rho_{xy}$(\textit{H}) measured at temperatures ranging from 0.35~K to 50~K in YbB$_{12}$ sample N1 up to 14~T. The low-field anomaly becomes weaker above 0.9~K and completely disappears at 2.5~K. The negative Hall signal at higher temperatures indicates that electron-like carriers dominate. (Inset)Magnetization $M$ of sample N1 versus magnetic field $H \parallel  [100]$ ranging from $-7~$T to $+7~$T at 1.8~K. The low-field $M(H)$ exhibits featureless and monotonic behavior. (b)~Expanded view of the $\rho_{xy}(H)$ curves measured at $T > 15~$K. (c)~Arrhenius plot: $\ln|\rho_{xy}|$ at $\mu_0H$ = 1~T plotted against $1/T$.
The linear fits in the temperature ranges $20~{\rm  K} \leq T \leq 50~$K and
$6~{\rm  K}\leq T \leq 12.5~$K yield Kondo gap widths $\Delta_1 = 18.2~$meV and $\Delta_2 = 3.2~$meV, respectively.
}
\label{FigHallPPMS}
\end{figure}

\begin{figure*}[htbp!]
\includegraphics[width=0.85\textwidth]{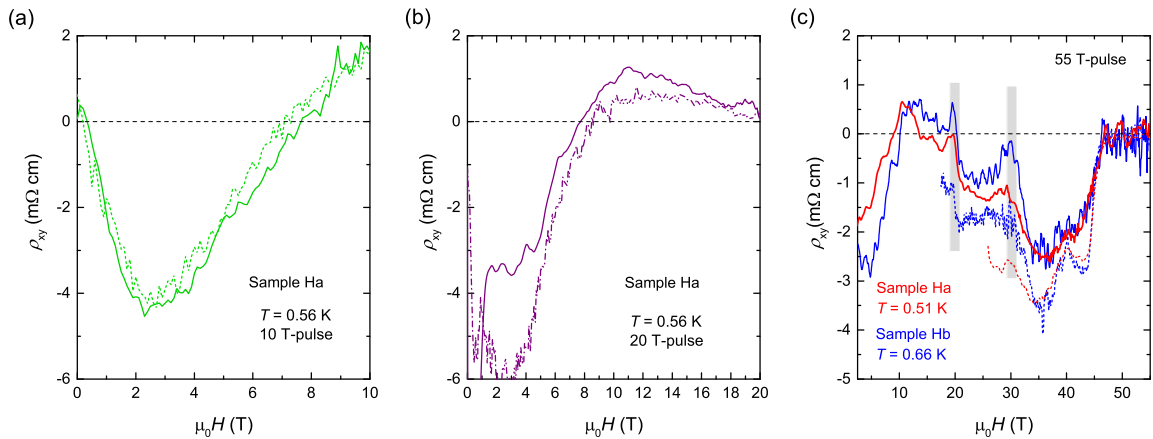}
\caption{Field dependence of Hall resistivity in YbB$_{12}$ measured in pulsed magnetic fields for (a)~sample Ha at 0.56~K up to 10~T, (b)~sample Ha at 0.56~K up to 20~T, (c) ~sample Ha at 0.51~K and sample Hb at 0.66~K, in magnetic fields of up to 55~T. Solid and short-dashed lines are field-upsweeps and downsweeps, respectively.
The $\rho_{xy}(H)$ curves are obtained by subtracting negative from positive field sweeps. The two light gray bars in (c)~mark the fields corresponding to possible Lifshitz transitions.
}
\label{FigHallPulsed}
\end{figure*}

The unusual low-field behaviour of the Hall effect ({\it i.e.,} the initially negative $\rho_{xy}$ with a minimum between 2 and 3\,T and sign change at $6-7~$T) is rapidly suppressed for $T \gtrsim 0.9~$K and completely disappears at $T$ = 2.5\,K [Fig.\,2(a)]. At higher $T$, $\rho_{xy}$ is negative with no sign change up to 14\,T, but is still not linear in $H$ [Fig.\,2(a)]. Such nonlinearity is most apparent at higher magnetic fields and gradually becomes weaker upon warming, whereas a quasi-$H$-linear behavior shows up at low $H$ [Fig.\,2(b)]. Values of $\ln|\rho_{xy}|$ at $\mu_0H = 1~$T are plotted against $T^{-1}$ in Fig.\,2(c); we extract the Kondo gap $\Delta$ by applying linear fits (via $\rho_{xy}(T)\propto n^{-1} \propto \exp(\Delta/2k_{\rm B} T)$, where $n$ is the carrier density and $k_{\rm B}$ the Boltzmann constant).
The fits suggest the presence of two different gaps, with $\Delta_1 = 18.2\pm 0.5~$meV between 50 and 17.5~K
and $\Delta_2 = 3.2 \pm 0.5~$meV between 6 and 15~K. Similar two-gap behavior has been observed in previous transport studies of YbB$_{12}$~\cite{Iga} and SmB$_6$~\cite{Shahrokhvand}; it is attributed to the complex level mixing in the hybridization process related to the crystal-field splitting of the $4f$ states~\cite{Shahrokhvand,SugiyamaHighField,Terashima2017JPSJ}. The values of $\Delta_1$ and $\Delta_2$ deduced here are slightly different from those obtained from $\rho_{xx}$~\cite{Xiang2018SdH} but are consistent with those from earlier Hall measurements~\cite{Iga}. Therefore, in strong contrast to the situation below $\approx 1~$K (Fig.\,1(a)), the low-field $\rho_{xy}$ for $T >6~$K seems to be dominated by the thermally-activated carrier density $n$ in YbB$_{12}$; it may be thought of as a reasonably conventional Hall effect with $\rho_{xy} = R_{\rm H} \mu_0H$, where $R_{\rm H} = (ne)^{-1}$. On the other hand, the unusual low-$H$ Hall behaviour for $T< 1.5~$K shows that a thermal activation description is insufficient to account for all of the transport properties of YbB$_{12}$.

\subsection{Origin of the Hall anomaly}
\label{chap03:2}

The low-field Hall anomaly is observed in three YbB$_{12}$ samples from two different batches: sample N1 (Fig.\,2(a)) and N2 (Fig.\,1(a)) from batch 1; sample Ha (Fig.\,3(a)) from batch 2. The behavior of $\rho_{xy}(H)$ in these samples shows qualitative consistency, suggesting that the anomaly is an intrinsic property of YbB$_{12}$. We now discuss possible mechanisms and examine their validity in the context of our Hall data.

The anomalous Hall effect (AHE) often results in unusual Hall-effect data that do not scale linearly with $H$, especially at low magnetic fields. In ferromagnets, the ``conventional" AHE is proportional to the magnetization $M$~\cite{AHE}. However, the $M(H)$ in YbB$_{12}$ (inset to Fig.~2(a)) is too small to result in a significant AHE. Moreover, it does not exhibit any form of unusual field dependence that could be responsible for the strong curvature of the Hall data. (Note that the small size of $M(H)$ in YbB$_{12}$ comes from the fact that it is a KI, so that the Yb local moments are fully screened in the ground state.)

Recently, an ``unconventional" AHE has been proposed for nonmagnetic topological materials; this originates from nonzero Berry curvature in the vicinity of Weyl nodes or gapped band crossings~\cite{ZrTe5,KV3Sb5}. For YbB$_{12}$, however, no such topological objects have been proposed for the band structure of YbB$_{12}$. Moreover, even if they dominate the electrical transport, topological surface states should not exhibit an AHE in the absence of magnetism.

In many heavy fermion compounds there is a third type of AHE that results from skew scattering: in magnetic fields, the conduction electrons are scattered asymmetrically by the magnetic moments of the
$f-$electrons~\cite{AHEMixedValence,FIBYbB12Sato}.
Though this effect can contribute a large portion of the measured Hall signal, this usually occurs at high $T$, where the localized $f$-electrons are not fully screened so that they can act as scattering centers~\cite{FertAHE}. At very low $T$, the skew scattering effect only involves residual impurities; consequently, the corresponding AHE becomes rather weak compared to the conventional Hall effect~\cite{YbRh2Si2AHE,YbRh2Si2Lifshitz,UTe2Meta}.
Moreover, most of the models of the AHE in heavy-fermion systems predict $R_{\rm H} \propto \chi$~\cite{YFYangAHE}, where $\chi = M/H$ is the magnetic susceptibility. This is nearly $H$-independent (inset of Fig.\,2(a)) in YbB$_{12}$. Therefore, skew-scattering-induced AHE can be excluded as the origin of the $H$-nonlinear Hall anomaly in YbB$_{12}$.

In the low-field limit ($\mu H \ll 1$, where $\mu$ is the carrier mobility), the Hall coefficient can be severely affected by details of the quasiparticle scattering; it thereby deviates from simple relationships such as {\it e.g.,} $R_{\rm H} = 1/ne$~\cite{Hurd_Hall}. It is well known that in 2D electron systems, such as the surface states in YbB$_{12}$, the amplitude and even the sign of the low-field Hall effect depends on the geometry (local curvature) of the FS~\cite{OngGeometry}. Nevertheless, ARPES data suggest a single Fermi contour at the center of the surface Brillouin zone with no apparent anisotropic curvature~\cite{ARPESYbB12}. Therefore, the FS geometry arguments are perhaps irrelevant here. Alternatively, anisotropic scattering can stem from antiferromagnetic (AFM) fluctuations with certain wave vectors~\cite{HallCeCoIn5,NFLCeCoIn5}. We remark that in YbB$_{12}$, the presence of dispersive in-gap spin excitons due to short-range AFM correlations has been confirmed~\cite{AFMfluc,SpinExciton}. In particular, the presumably reduced strength of Kondo coupling at the surface can give rise to enhanced AFM stabilities and even push the system towards quantum criticality \cite{SmB6SpinExcitation}; spin-exciton-mediated scattering naturally provides strong anisotropic scattering channels for the surface quasiparticle states.

While it seems that most of the established models do not account for the Hall anomaly in YbB$_{12}$, we stress that this feature can be almost perfectly fitted by an empirical expression (yellow line in Fig.\,1(b)):
\begin{equation}
R_{\rm H} = R_{\rm H}^0 + A_{\rm H}\exp[-\eta(\mu_0H)^{\frac{3}{2}}].
\label{empirical}
\end{equation}
Here, $R_{\rm H}^0$ is a field-independent term with the value ${\rm 0.3~m\Omega cm T^{-1}}$, and $A_{\rm H} \exp[-\eta(\mu_0H)^{\frac{3}{2}}]$ is the ``anomalous" term that gives rise to the nonlinear behaviour, with $A_{\rm H}$ and $\eta$ being fitting parameters. Significantly, a similar Hall enhancement at low $H$ has been observed in the ``strange-metal'' phase of iron-based superconductors BaFe$_2$(As$_{1-x}$P$_x$)$_2$~\cite{HallBaFeAsP} and FeSe$_{1-x}$S$_x$~\cite{HallFeSeS}. In both cases, the  enhancement is also $\propto \exp[-\eta(\mu_0H)^{\beta}]$ with the exponents $\beta = 1$ for BaFe$_2$(As$_{1-x}$P$_x$)$_2$~\cite{HallBaFeAsP} and $\beta = 2$ for FeSe$_{1-x}$S$_x$~\cite{HallFeSeS}.
In both iron-based superconductors, the strange-metal state is related to proximity to a putative quantum-critical point (QCP)  (a pure nematic QCP in FeSe$_{1-x}$S$_x$ \cite{HallFeSeS,NFLFeSeS} and a (mainly) AFM QCP in BaFe$_2$(As$_{1-x}$P$_x$)$_2$~\cite{ShigeruBaFeAsP,ShibauchiARCMP}).
Hence, the successful fit to Eq.\,\ref{empirical} with $\beta = 3/2$ strongly suggests a strange-metal state in YbB$_{12}$ and it would be of interest to explore whether critical magnetic instabilities, possibly induced by the weakened Kondo coupling at the surface \cite{SmB6SpinExcitation}, contribute to the formation of such states.

The two terms in Eq.\,\ref{empirical} represent a strange-metal contribution $A_{\rm H} \exp[-\eta(\mu_0H)^{\frac{3}{2}}]$, plus a constant, $R_{\rm H}^0$. If one for a moment assumes $R_{\rm H}^0$ to be the ``conventional'' part of the Hall effect, it corresponds to a rather high carrier density $n \simeq 2.08 \times 10^{18}~{\rm cm}^{-3}$ (equivalent to $n^{\rm 2D} \simeq 4.4\times 10^{16}~{\rm cm}^{-2}$).
This means $R_{\rm H}^0$ cannot be interpreted as due to a conventional band contribution. Such a heavily modified apparent $n$ could be the consequence of emergent quasiparticle decoherence, as reported in the strange-metal regime of cuprate high-$T_c$ superconductors~\cite{PutzkeHall}. Thus, it seems that the low-$T$, low-$H$ $R_{\rm H}$ in YbB$_{12}$ dominated entirely by strange-metal physics.
Since the low-$T$, low-$H$ electrical conduction in YbB$_{12}$ is thought to be due to the surface channel \cite{FIBYbB12Sato}, the strange metal state likely resides on the surface, where the quasiparticles lose coherence through scattering from abundant bulk excitations. We will further discuss this in Sec.\,\ref{chap06}.

\section{Evidence for field-induced Lifshitz transitions}
\label{chap04}

\subsection{Hall resistivity at high magnetic field}
\label{chap04:1}

\begin{figure}[htbp!]
\centering
\includegraphics[width=0.46\textwidth]{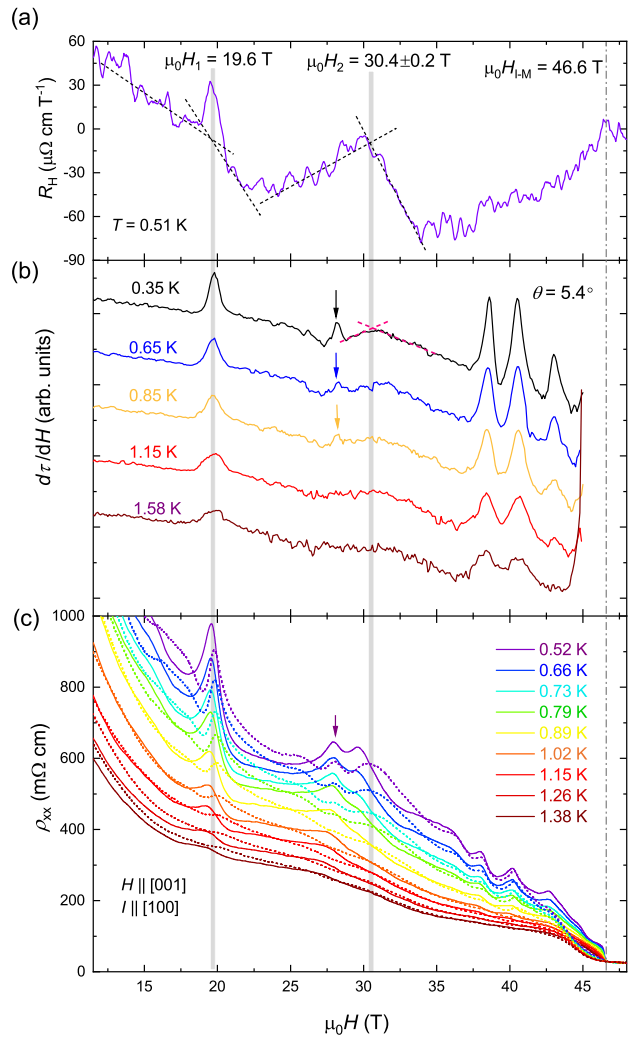}
\caption{Comparison of the field dependence of (a)~Hall coefficient $R_{\rm H}$, (b)~field derivative of the magnetic torque, d$\tau$/d$H$, taken with tilt angle $\theta = 5.4^\circ$ between $H$ and the [001] direction and (c)~transverse resistivity $\rho_{xx}$ versus $H$ applied along [001] in YbB$_{12}$. The insulator-to-metal transition field $\mu_0H_{\textrm{I-M}}$ is 46.6~T in this sample, as denoted by the vertical dash-dot line. Clear dHvA and SdH oscillations appear below $H_{\textrm{I-M}}$ as shown in (b) and (c), respectively. The vertical gray thick lines represent two characteristic fields: a nearly $T$-independent signature $\mu_0H_1 = 19.6~$T manifested by (a) a slope change in $R_H$ and peaks in (b) d$\tau$/d$H$ and (c) $\rho_{xx}$($H$), and a second signature occurring at $\mu_0H_2 = 30.4\pm0.2~$T which is implied by the kinks in $R_H$ and d$\tau$/d$H$ (dashed lines in (a) and (b), respectively). This latter feature roughly corresponds to a local maximum in MR for the downsweeps. Alternatively, the small cusp in d$\tau$/d$H$ at approximately 28\,T (arrows) is identified as an extremum of quantum oscillations (see Fig.\,12 and Appendix Sec.\,\ref{chap07:4}).
}
\label{FigLifshitz}
\end{figure}

With increasing $H$, the electrical transport in YbB$_{12}$ exhibits a crossover from the low-field surface-dominated regime to the high-field bulk-conducting regime~\cite{FIBYbB12Sato} where quantum oscillations of 3D unconventional, charge-neutral fermions have been detected~\cite{Xiang2018SdH}. The behavior of $\rho_{xy}(H)$ is in agreement with this picture. As displayed in Fig.\,3(b), at $T = 0.56$\,K, the positive $\rho_{xy}$ developed above the sign change around 7\,T (Fig.\,3(a)) is slowly suppressed, leading to a second sign change from positive to negative at $\approx 20$\,T. For $\mu_0 H \gtrsim 20$~T, $\rho_{xy}$ appears to be electron-like all the way up to the I-M transition field $H_\textrm{{I-M}}$ [$\mu_0H_\textrm{{I-M}} = 46.6$\,T for samples Ha and Hb, Fig.\,3(c)]; this is identified as the contribution of bulk carriers. The minimum ({\it i.e.,} the largest magnitude negative value) $\rho_{xy} \approx -2.5~{\rm m}\Omega$cm occurs at $\approx 35$\,T [Fig.\,3(c)]. In the field-induced KM state $(H >H_\textrm{{I-M}})$~\cite{Terashima2018PRL} the Hall signal drops below our experimental resolution.

The measurement of tiny Hall signals in pulsed magnetic fields is affected by issues such as eddy-current heating (due to the large field-sweeping rate), the magnetocaloric effect and induced-voltage pickup in the circuit.
In the present case, difficulties may arise because the longitudinal resistivity in YbB$_{12}$ is both large ($\rho_{xx}/\rho_{xy} >100$ over the entire field range) and very $T$-dependent~\cite{Xiang2018SdH}.
Hence, an exact, quantitative extraction of the Hall voltage using a comparison of positive and negative field sweeps depends on $T$ being identical throughout both pulses. A slight mismatch in $T$ is likely responsible for the difference between the data recorded on the up- and downsweeps of the field in Fig.\,3(c).
Whilst the magnitudes of the $\rho_{xy}$ signals shown in Fig.\,3(c) should therefore be treated with caution, it is clear that two noticeable signatures in the high-field $\rho_{xy}$ curves are very consistently observed and therefore very likely to be intrinsic.
These are two sharp slope changes at $\approx 20\,$T ({\it i.e.,} near the second sign change) and $\approx30\,$T marked by gray vertical bars in Fig.\,3(c).
As discussed in detail in the following sections, we propose that these features are likely to signify field-induced Lifshitz transitions, {\it i.e.,} electronic topological transitions that describe changes in the FS topology \cite{LifshitzJETP}.

\subsection{Comparison with the features in magnetic torque and magnetoresistance}
\label{chap04:2}

\begin{figure*}[htbp!]
\includegraphics[width=0.85\textwidth]{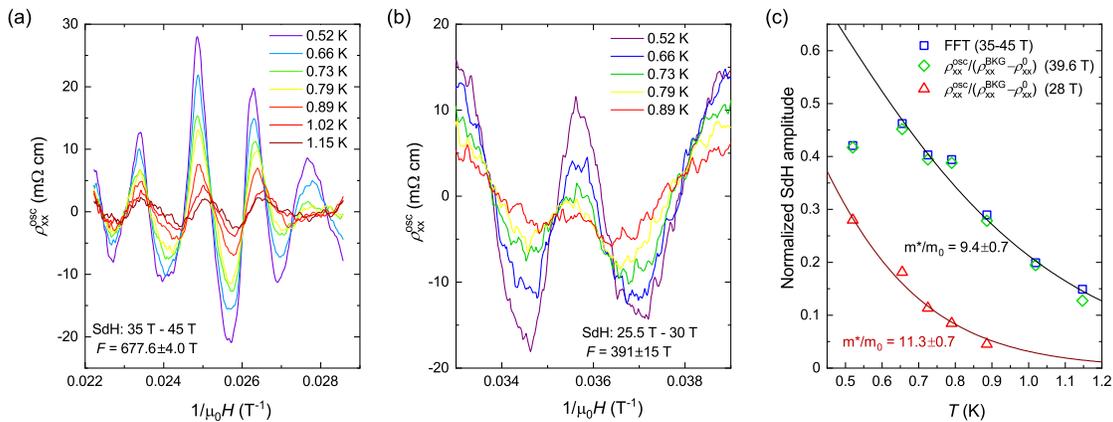}
\caption{The amplitude of the oscillatory component of resistivity, $\rho^{osc}_{xx}$, at various temperatures in the field ranges (a)~$35 \leq \mu_0H \leq 45~$T and (b)~$25.5 \leq \mu_0H \leq 30~$T. $\rho^{osc}_{xx}$ is obtained by background subtraction of a polynomial from the $\rho_{xx}(H)$ curves shown in Fig.\,4(c). The oscillation frequencies are mentioned in each panel. Note that for the field range displayed in (a), the oscillations have a single frequency which is supported by the dHvA measurements (see Supplemental Material Fig.\,1(a) and (b), \cite{Supplement}).
(c)~Lifshitz-Kosevich fits of the $T-$dependence of $\rho^{osc}_{xx}$ shown in (a) (green diamonds) and (b) (red triangles) normalized by the power-law $T$-dependent background resistivity (see Sec.\,\ref{chap05:3} for details). The temperature dependence of the FFT peak height (the FFT is performed after
normalizing $\rho^{osc}_{xx}$ by the background resistivity) of the oscillations shown in (a) (blue squares) is consistent with that of normalized $\rho^{osc}_{xx}$. All data were measured on sample Ha.
}
\label{FigSdH}
\end{figure*}

Figure\,4 shows the field dependence of the Hall coefficient $R_{\rm H}$ (Fig.\,4(a)), the derivative of the magnetic torque d$\tau/$d$H$ [Fig.\,4(b)] and $\rho_{xx}(H)$ [Fig.\,4(c)] measured in YbB$_{12}$ sample Ha. Several signatures show up in all three quantities. At $\mu_0H_1 = 19.6\,$T, a kink (slope change) is observed in $R_{\rm H}$ [indicated by the dashed lines in Fig.\,4(a))]. (Note that the peak at $H \approx H_1$ may be caused by a slightly incomplete cancellation of the picked-up $\rho_{xx}$ terms between the negative and positive field sweeps due to small temperature difference.) At the same field, strong peaks occurs in both d$\tau/$d$H$ and $\rho_{xx}$. The features at $H_1$ gradually smear out at higher $T$, yet the characteristic field $H_1$ is almost $T$-independent. At around 28\,T, a small cusp appears in d$\tau/$d$H$ [arrow in Fig.\,4(b)]. It approximately aligns with a local maximum in both field-upsweeps and downsweeps of MR at temperatures below $\sim$1\,K [Fig.\,4(c)]. We attribute this feature to an extremum of the quantum oscillation occurring between 25.5\,T and 30\,T (Appendix Sec.\,\ref{chap07:4}), which will be further introduced in Sec.\,\ref{chap05:1}. Both $R_{\rm H}$ and d$\tau/$d$H$ show another kink [crossing of the two dashed lines in Fig.\,4(a) and (b), respectively] at $\simeq$30.4\,T with the uncertainty of about 0.2\,T; this feature, identified as a second characteristic field $H_2$, is in line with a rounded maximum in the downsweeps of MR. All the signatures mentioned above were reproduced in the measurements of sample N1 (from a different batch), wherein the characteristic field $H_1$ stays unchanged but $\mu_0H_2$ = 31.3\,T is slightly higher (Appendix Sec.\,\ref{chap07:4}).

Some of the signatures we present here were explained as field-induced metamagnetic transitions in our earlier work~\cite{Xiang2018SdH}. These attributions followed theoretical predictions that an AFM state will appear in KIs subjected to large magnetic fields that are below the gap closure~\cite{BeachAFM,OhashiKI}.
However, it was later shown that the corresponding features are {\it not} observed in thermodynamic quantities such as the magnetocaloric effect~\cite{Terashima2018PRL}, magnetostriction and $M$ \cite{Xiang2021pulsed}. In particular, the absence of a signature in $M$ strongly suggests that they are not metamagnetic transition fields.

Instead, the signatures at the characteristic fields in YbB$_{12}$ closely resemble those in the heavy-fermion metal CeIrIn$_5$, where kinks in $\tau$ can be resolved but no feature appears in $M$. The latter kinks are interpreted as field-induced Lifshitz transitions that are not accompanied by metamagnetic transitions~\cite{AokiCeIrIn5}: the modification in FS topology causes a change in the transverse magnetization (which is captured in $\tau$) but has very weak impact on the longitudinal magnetization $M$. Although in numerous heavy fermion materials the Zeeman-driven Lifshitz transition does correspond to a simultaneous metamagnetic transition \cite{MetaMag}, this does not happen in CeIrIn$_5$ due to the robustness of Kondo physics in this material against magnetic fields~\cite{AokiCeIrIn5}. Such an argument will also be valid in YbB$_{12}$, where the field-suppression of the Kondo gap cannot be understood as the destruction of Kondo correlation~\cite{Terashima2018PRL,Xiang2021pulsed}. Consequently, we assign the characteristic fields shown in Fig.\,4 and Fig.\,12 to putative Lifshitz transitions (here we note that, whereas these transitions are likely to correspond to changes in FS topology, we do not have strong evidence that they are of the order$-2\frac{1}{2}$ nature as for the rigorously defined Lifshihtz transitions \cite{LifshitzJETP}).

Further evidence for the possible Lifshitz transitions comes from the transport measurements: it is well known that Lifshitz transitions often yield kinks in the Hall effect and MR~\cite{YbRh2Si2Lifshitz,PourretYbRh2Si2,PfauYbNi4P2}. Whereas Lifshitz transitions may cause small changes in the field evolution of the density of states (DOS), which produce only weak signatures in thermodynamic quantities, they can correspond to the appearance or vanishing of scattering channels, strongly modifying the transport behaviors. Considering two basic type of Lifshitz transitions, namely the FS ``neck breaking" and ``void creation" transitions, a simple model predicts that they are manifested by local maxima and minima of the electrical resistivity, respectively~\cite{PfauYbNi4P2}. Following this idea, all of the putative Lifshitz transitions in YbB$_{12}$ are presumably of the ``neck breaking" type. In particular, the feature observed at $H_2$ may be interpreted as a Lifshitz transition that barely changes the DOS and is thus not obvious in $\tau$ [Fig.\,4(b)]. Further information about the quasiparticle scattering is needed to pin down the details of the Lifshitz transitions. In this context, thermopower measurements under high magnetic fields~\cite{PourretYbRh2Si2} may be of great importance.

Perhaps the most direct verification of a Lifshitz transition is a discontinuity in the quantum-oscillation frequency, as has been observed in {\it e.g.,} CeIrIn$_5$~\cite{AokiCeIrIn5}, YbRh$_2$Si$_2$~\cite{RourkeYbRh2Si2} and CeRu$_2$Si$_2$~\cite{CeRu2Si2dHvA}. In Sec.\,\ref{chap05:1} we will show that the SdH frequency changes drastically within the approximate field range $30 - 33\,$T. This strongly supports the occurrence of a Lifshitz transition in the corresponding field window and it is tempting to assign it to the characteristic field $H_2$.

Given the nonmetallic transport behavior for $H < H_{\textrm{I-M}}$ in YbB$_{12}$, an important question is ``what kind of Fermi surface plays host to the Lifshitz transitions?'' As we will discuss further in Sec.\,\ref{chap06}, a reasonable hypothesis is that the Lifshitz transitions occur in the FS of charge-neutral quasiparticles~\cite{Xiang2021pulsed} and show up in the charge transport via interband scattering between the neutral quasiparticles and any conventional charge-carrying electrons. At this point, without knowing the exact form of the neutral quasiparticle bands, we cannot provide a detailed scenario for the possible Lifshitz transitions. However, it is likely that they involve abrupt modifications of the neutral FSs that are triggered by the field-induced gap narrowing (see Sec.\,\ref{chap06:4}).

\section{Quantum oscillations in the Kondo insulator state}
\label{chap05}

\subsection{SdH effect in the field range 25.5\,T$\leq \mu_0H \leq$ 30\,T}
\label{chap05:1}

We previously reported dHvA and SdH oscillations in the KI state ($H < H_{\textrm{I-M}}$) of YbB$_{12}$ at magnetic fields $\mu_0H \gtrsim 33-35~$T~\cite{Xiang2018SdH,Sato2019thermal}. Here we show that SdH oscillations actually emerge at lower $H$. As plotted in Fig.\,4(c), $\rho_{xx}(H)$ traces recorded on the  downsweeps of $H$ (short-dashed lines) exhibit a double-dip feature in the field range $25.5~{\rm T}\leq \mu_0H \leq 30~$T. This feature is most evident at low $T$, gradually fading upon warming. While it is relatively clearer in sample Ha [Fig.\,4(c)], this feature is also present in other batches of samples [see the shaded region in the inset of Fig.\,12(c) and Appendix Sec.\,\ref{chap07:4}]; it was simply overlooked in our earlier studies.

Figure 5(a) and (b) show the oscillatory component of $\rho_{xx}$, $\rho_{xx}^{osc}$, obtained  using polynomial background subtraction for ${\rm 35~T} \leq \mu_0H \leq 45~$T and ${\rm 25.5~T}\leq \mu_0H \leq 30~$T respectively. For both field ranges, the amplitudes of the oscillations shrink with increasing $T$ in a similar way. In Fig.\,5(c) we plot the $T-$dependence of the normalized amplitude of the SdH effect and fit it using the Lifzhitz-Kosevich (LK) formula \cite{shoenberg2009magnetic}:
\begin{equation}
\rho_{xx}^{osc}(T) \propto \frac{\frac{2\pi^2 k_{\rm B} m^* T}{e\hbar \mu_0H}}{\sinh(\frac{2\pi^2 k_{\rm B} m^* T}{e\hbar \mu_0H})},
\label{LK_temp}
\end{equation}
where $m^*$ is the cyclotron mass for motion in the plane perpendicular to the applied $H$. (We note that to obtain reliable cyclotron masses, $\rho_{xx}^{osc}$ must be normalized by the background resistivity before applying the LK fits; see Sec.\,\ref{chap05:3}.)

It appears that the LK formula [solid lines in Fig.\,5(c)] describes the oscillation amplitudes well except for the lowest temperature point for the ${\rm 35~T} \leq \mu_0H \leq 45~$T data (this point is affected by the change of background resistivity behavior for $T\lesssim 0.5~$K).
Interestingly, while the fast-Fourier transform (FFT) gives SdH frequencies $F = 678 \pm 4.0~$T and $391 \pm15~$T for ${\rm 35~T}\leq \mu_0H \leq 45~$T and ${\rm 25.5~T} \leq \mu_0H \leq 30~$T, respectively, the LK fits yield comparable cyclotron masses $(9.4m_0$ and $11.3m_0$, respectively, where $m_0$ is the mass of the free electron) for the two field regions. The similarity of the masses excludes magnetic breakdown as an explanation for the frequency change, which predicts that the mass of the large breakdown orbits equals the sum of the masses of the involved small orbits (based on the relation $m^* \propto (\partial S/\partial \epsilon)\mid_{\epsilon = E_F}$, where $S$, $\epsilon$ and $E_F$ are the orbit area, the energy and the Fermi level, respectively)\,\cite{shoenberg2009magnetic}. The consistency of the mass instead suggests that all of the SdH oscillations may originate from the same quasiparticle band(s), whilst a Lifshitz transition between the two field regions changes the extremal orbit area. This Lifshitz transition may be associated with the characteristic field $H_2$ in Fig.\,4. Unfortunately, the SdH effect occurring below $\approx 31~$T is only seen in the downsweep MR curves, and no more than one and a half periods are resolved. The hysteresis between up- and down-sweeps of $H$ suggests that these SdH oscillations may be  associated with ``metastable" states that are induced by, and/or only persist over the ms timescales of the pulsed magnetic fields.

\begin{figure}[htbp!]
\centering
\includegraphics[width=0.41\textwidth]{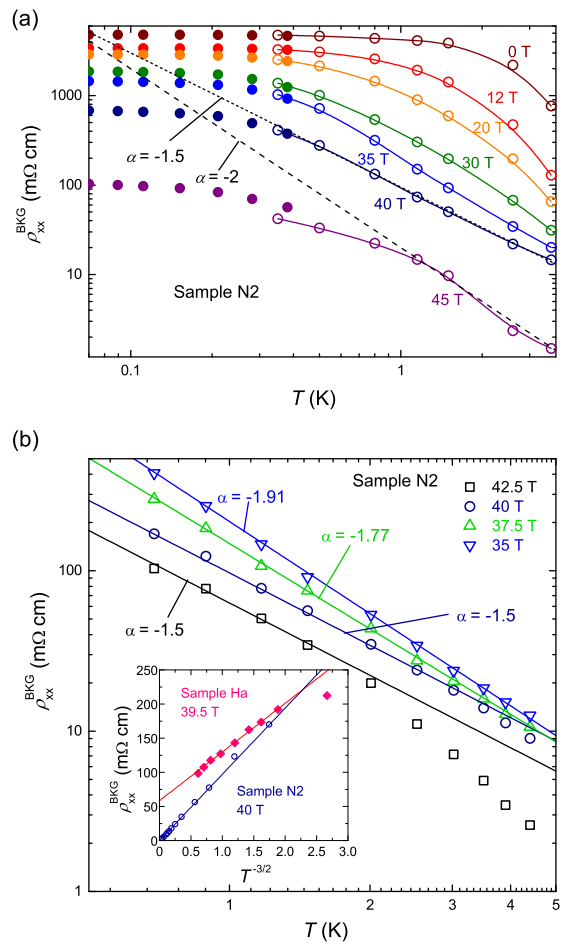}
\caption{(a) The nonoscillatory background resistivity $\rho_{xx}^{\rm BKG}$ of YbB$_{12}$ sample N2 as a function of temperature at several magnetic fields from 0 to 45~T, plotted on a log-log scale. Open and solid symbols are data taken in the $^3$He cryostat at magnetic field tilt angle $\theta = 7.4^\circ$
and in the dilution refrigerator at $\theta = 8.5^\circ$ in the 45~T hybrid magnet. Solid lines are guides to the eye. Dashed and dotted lines denote $\rho_{xx} \propto T^\alpha$ with $\alpha = -2$ and $\alpha = -1.5$, respectively. (b) $\rho_{xx}^{\rm BKG}(T)$ of sample N2 at $\mu_0H = 35~$T (down triangles), 37.5~T (up triangles), 40~T (circles) and 42.5~T (squares), measured in a pulsed magnet with $H$ applied along [001]. Solid lines indicate the power-law $T$-dependence $\rho_{xx}^{\rm BKG}(T) = A(H)T^{\alpha}$. (Inset)$\rho_{xx}^{\rm BKG}$ plotted against $T^{-3/2}$ for sample N2 ($\mu_0H= 40~$T) and Ha ($\mu_0H = 39.5~$T).
}
\label{FigRT40T}
\end{figure}

\subsection{Nonmetallic transport at high fields}
\label{chap05:2}

To further understand the SdH effect observed in the KI state of YbB$_{12}$, we investigate the $T$-dependence of the nonoscillatory resistive background $\rho_{xx}^{\rm BKG}$ ({\it i.e.,} the $H$-polynomial term which is subtracted from the raw MR data to get $\rho_{xx}^{\rm osc}$)  over the field range where the SdH oscillations are apparent ($35-45\,$T). As displayed in Fig.\,6(a) and (b), over a field-dependent $T-$range $\rho_{xx}^{\rm BKG}$ in sample N2 follows an unusual $T^{\alpha}$ behavior, with the exponent $\alpha < 0$ between 35 and 45\,T. At 35\,T and 37.5\,T, $\rho_{xx}^{\rm BKG}$ is well described by $T^{\alpha}$ functions for $0.5 \lesssim T \lesssim 5\,$K with $\alpha$ = -1.91 and -1.77, respectively [Fig.\,6(b)]. For $\mu_0H = 40\,$T,
and $0.35 \lesssim T  \lesssim 3$~K, $\rho_{xx}$ shows a perfect $T^{-1.5}$ dependence. On increasing the field to 42.5\,T, $\alpha$ = -1.5 is still valid but only for $T \lesssim 2\,$K [Fig.\,6(b)]. At 45\,T ({\it i.e.,} close to the I-M transition) a $T^{-2}$ behavior emerges over a narrow $T$ range above 1\,K [Fig.\,6(a)].

Whilst the background resistivities in samples N2 and Ha can be fitted to the simple relationship
$\rho_{xx}^{\rm BKG}(T) = \rho_{xx}^{0} +A(H)T^{\alpha}$,
the fitting parameters $\rho_{xx}^{0}$ and $A$ are sample-dependent.
In particular, $\rho_{xx}^{0}$ is infinitesimal for sample N2 yet takes a value $\approx 60~{\rm m \Omega cm}$ in sample Ha [inset of Fig.\,6(b)].

The $T^{\alpha}$ behavior with $-2.0 \lesssim \alpha \lesssim -1.5$ is strong evidence for nonmetallic charge transport upon which the SdH effect emerges. This behavior contrasts sharply with that of conventional SdH oscillations associated with a FL state. The underlying physics leading to the negative $\alpha$ is elusive at present. Theoretically, a $\rho \propto T^{-\frac{3}{2}}$ dependence can be attributed to ionized impurity scattering in non-degenerate semiconductors \cite{ionizedimp} or electron-phonon-dominated inelastic scattering in heavily disordered systems~\cite{WeakLocalization}. Neither electronic system is expected to produce quantum oscillations on its own.

More unusually, recent works on electrical transport in topological semimetals with small Fermi energies and charged impurities also predict $\rho \propto T^{\alpha}$ (with $\alpha <0)$~\cite{WeylScreening,DiracScreening,WeylDisorderMR}. Indeed, it has been suggested that in KIs, impurity scattering leads to non-Hermitian in-gap states that effectively close the Kondo gap and turn the system into a nodal semimetal~\cite{nonHermitian,HarrisonNode}. Although we later show that this does not happen in YbB$_{12}$ at $H=0$ (see Appendix Sec.\,\ref{chap07:1}), it is not clear whether it is relevant at high $H$, where the gap width is reduced. Whereas a $T^{-1}$ dependence is predicted when only the thermal excitation of carriers across the nodal cone is taken into account~\cite{HarrisonNode}, $\rho \propto T^{-4}$ is obtained if the $T$-dependence of the scattering rate and thermal screening of impurities are included~\cite{WeylScreening,DiracScreening}. The exponent $-2.0 \lesssim \alpha \lesssim -1.5$  may therefore represent vestiges of these effects; for instance, the putative $T^{-4}$ behavior is
predicted to evolve~\cite{DiracScreening} into a weakly-$T$-dependent ``residual" resistivity at low $T$, qualitatively consistent with our observations for $T \lesssim 0.3\,$K [Fig.\,6(a)]. However, in proposed Weyl semimetals with small Fermi energies such as the iridate pyrochlores \cite{R2Ir2O7,Y2Ir2O7Weyl}, a low$-T$ saturation of $\rho$ is experimentally absent.

\subsection{Correction of the SdH amplitudes}
\label{chap05:3}

\begin{figure}[htbp!]
\centering
\includegraphics[width=0.46\textwidth]{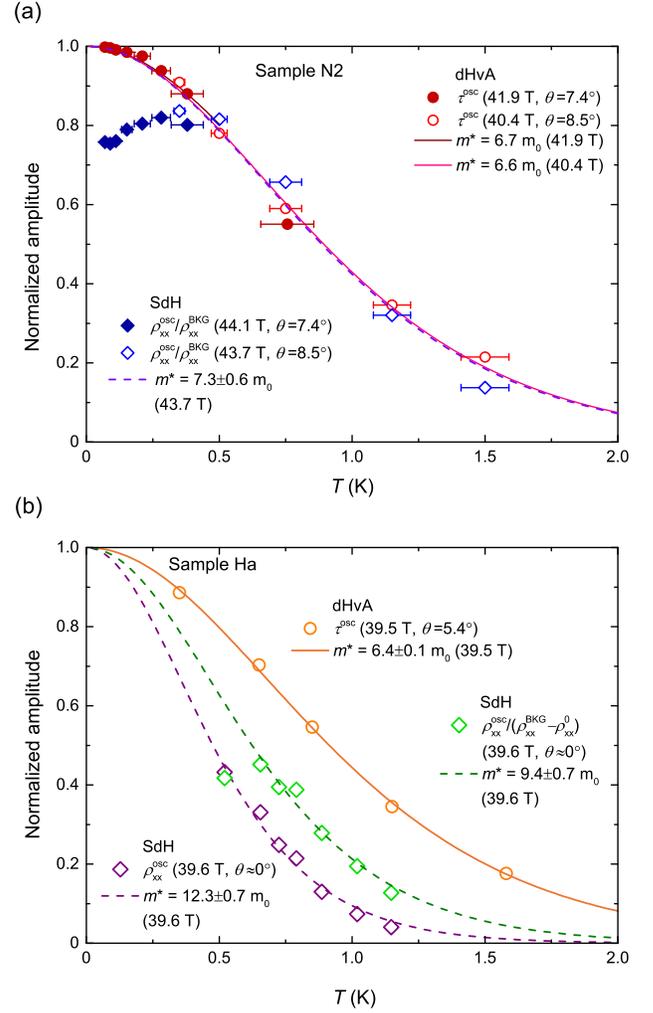}
\caption{Normalized quantum oscillation amplitudes versus temperature for (a) sample N2 and (b) sample Ha. The dHvA amplitude $\tau^{\rm osc}$ is obtained by subtracting a polynomial background from the magnetic torque $\tau$($H$) \cite{Xiang2018SdH}. For sample N2, the SdH amplitude is taken as $\rho^{\rm osc}_{xx}$/$\rho^{\rm BKG}_{xx}$, i.e., the oscillatory part of resistivity $\rho^{\rm osc}_{xx}$ divided by the non-oscillatory background at the corresponding magnetic field. For sample Ha, a $T$-independent component $\rho^0_{xx}$ is further removed from the non-oscillatory resistivity background $\rho^{\rm BKG}_{xx}$ (see text). The absolute $\rho^{\rm osc}_{xx}$ without normalization using the background resistivity (purple diamonds) is also shown for comparison. Solid and dashed lines are fits to the Lifshitz-Kosevich formula for dHvA and SdH amplitudes, respectively. The inferred effective masses are shown besides the fits. Note that all the nonmonotonic $T$-dependence of the SdH amplitudes are caused by the normalization, see the raw data presented in Supplemental Material Fig.\,1(c), \cite{Supplement}.
}
\label{FigLK}
\end{figure}

Earlier works describing quantum oscillations in the KI state of YbB$_{12}$ noted an apparent contradiction; the cyclotron mass determined from the SdH effect $(14-15 m_0)$ was much heavier than that obtained from dHvA oscillations $(6.5-7 m_0)$~\cite{Xiang2018SdH}.
The discrepancy may now be understood as follows.
Whereas the thermal damping of the dHvA oscillations is due only to phase smearing caused by the $T$-dependent broadening of the Fermi-Dirac distribution function (LK formula, Eq.\,\ref{LK_temp}), SdH oscillations in electrical transport represent an additional convolution of this effect with the background resistivity. The latter will be affected by  $T$-dependent scattering rates, an effect parameterized via $\nu \simeq \zeta(T) \tilde{g}$~\cite{Pippard}, where $\nu$ is the scattering rate (inverse relaxation time), $\tilde{g}$ is the ($T-$independent) DOS and  $\zeta(T)$ describes the $T-$dependent scattering processes. It is clear that $\zeta(T)$ should contribute equally to the oscillatory and non-oscillatory (background) components of the resistivity, so that $\rho^{\rm osc}(T)$ must be normalized by dividing by $\rho^{\rm BKG}(T)$ before fitting to the LK formula~\cite{AdamHolstein,RichardsMg}.

In our earlier work, $\rho_{xx}^{\rm osc}$($T$) was normalized in the traditional manner~\cite{Pippard} using the zero-field resistivity $\rho_{xx}(T)$ \cite{Xiang2018SdH}. However, the distinctive $H$ and $T$ dependence of $\rho_{xx}$ (Fig.\,6 and Fig.\,9) suggests that $\zeta(T)$ at high $H$ differs considerably from that at $H=0$ ({\it i.e.,} it is also a function of $H$), so that a different normalization strategy must be employed.
Thus, the {\it apparent} discrepancy between the effective masses derived from SdH and dHvA effects is mainly attributable to $\zeta(H,T)$. Fig.\,7(a) shows a comparison for sample N2 of dHvA amplitudes and SdH amplitudes divided by  $\rho_{xx}^{\rm BKG}$ at the same $H$. LK fits to the data for $T>0.5\,$K (i.e., where $\rho_{xx}^{\rm BKG} \propto T^{\alpha}$) yield cyclotron masses from dHvA and SdH effects that are consistent within experimental uncertainties. This, and the fact that both sets of oscillations have the same frequency, strongly suggests that the same band of coherent quasiparticles (with an effective mass $\approx 7 m_0$) is responsible for both effects in YbB$_{12}$. Our recent pulsed-field studies reveal that these fermions also contribute quantum oscillations in the high-field KM state of YbB$_{12}$ \cite{Xiang2021pulsed}, acquiring additional features such as Zeeman splitting and a field-dependent SdH frequency.

The consistency between $\tau^{\rm osc}$($T$) and $\rho_{xx}^{\rm osc}$/$\rho_{xx}^{\rm BKG}$ breaks down below 0.5\,K [Fig.\,7(a)] as the latter exhibits nonmonotonic behavior, which is likely to be an artefact. (Note that this nonmonotonic $T$-dependence of
the SdH amplitudes results entirely from the normalization to the background resistivity, since the raw data are free from such behavior; see Supplemental Material \cite{Supplement}).
Moreover, the mass consistency is not as good for sample Ha, in which the $\rho_{xx}^{\rm osc}$ normalized to $\rho_{xx}^{\rm BKG}-\rho_{xx}^{0}$ [i.e., the $T^{\alpha}$ term in the inset of Fig.\,6(b)] still gives a heavier mass $(m^*/m_0 = 9.4\pm 0.7)$ than that from the dHvA effect $(m^*/m_0 =6.4\pm 0.1)$, as shown in Fig.\,7(b). Nevertheless, the value of SdH mass becomes closer to the dHvA mass after the normalization; by contrast, the LK fit of the absolute SdH amplitudes provides $m^*/m_0 =12.3\pm 0.7$ [Fig.\,7(b)].
Despite the additional complexity of the $T-$dependent background resistivity, the LK description is valid for the dHvA effect observed between 35 and 45\,T [Fig.\,7(a) and (b)] as well as the normalized SdH effect for 25.5\,T$\leq \mu_0H \leq$ 30\,T [Fig.\,5(c)]. As the LK formula is closely related to the derivative of the Fermi-Dirac distribution function, this is an unambiguous sign that the quantum oscillations in both field ranges are produced by fermions.

\section{Discussion}
\label{chap06}
\subsection{The two-fluid scenario}
\label{chap06:1}

The transport properties of YbB$_{12}$ in its Kondo-insulator state point to the presence of both
a Fermi surface of coherent quasiparticles - which, however, carry little or no charge - {\it and}  electrically
active charge carriers - some or all of which never gain full quasiparticle coherence - that are characteristic of a complex electrical {\it insulator}.
This evidence may be summarized as follows.\\

\noindent
{\it Fermi surface of coherent quasiparticles:}
\begin{enumerate}
\item
Quantum oscillations (both dHvA and SdH effects) are detected, with behaviors that are in agreement with the expectations for a 3D Fermi liquid.
\item
There is a cascade of signatures in the magnetic torque, Hall effect and magnetoresistance strongly suggesting field-induced Lifshitz transitions that change the topology of the Fermi surface.
\item
A finite, residual $T$-linear thermal conductivity,  a clear hallmark of itinerant gapless fermions, is observed as $T\rightarrow 0$~\cite{Sato2019thermal}.
\end{enumerate}

\noindent
{\it Complex electrical insulator:}
\begin{enumerate}
\setcounter{enumi}{3}
\item
Both zero-field and high-field resistivities exhibit characteristic nonmetallic behavior; at high magnetic fields
this is described using a $T^{\alpha}~ (\alpha <0)$ dependence.
\item
The absolute value of the resistivity is much larger than that of conventional metals~\cite{Xiang2018SdH}.
\item
Apart from brief excursions due to the SdH oscillations, the underlying MR is negative for all fields up to the I-M transition.
\item
The Hall coefficient shows a low-field anomaly that cannot be explained by the conventional Drude model for metals. However, the form of the Hall coefficient  is very similar to that of incoherent
quasiparticles in the strange-metal phase regime of high-$T_c$ superconductors.
\end{enumerate}
In Ref.\,\cite{Xiang2021pulsed} we used a two-fluid model
to give a phenomenological description of YbB$_{12}$ in both its KI and
KM states. In the latter state, the fluids are (i)~the conventional (heavy-fermion) charge carriers
liberated by the collapse of the Kondo gap at high fields
and (ii) a Fermi-surface of charge-neutral quasiparticles.
In the KI state, (i) a complex system of charge-carrying electrons and holes containing both thermally-activated carriers and other charged in-gap states (see Sec.\,\ref{chap06:2})
coexists with (ii) the same charge-neutral quasiparticles.
Under magnetic fields, the charge-neutral quasiparticles develop Landau levels
that lead to an oscillatory DOS. In both KI and KM states, these oscillations in the DOS result in the SdH effect in the resistivity
due to strong scattering between the two fluids (at a rate that is determined by the joint DOS and hence the Landau quantization).

Taken alongside other recent experimental observations, the Hall-effect data in this paper permit a better description of the involved system of charge-carrying electrons and holes in the KI state than has previously been possible. We therefore first turn to the likely origin of these carriers.

\subsection{Charge-carrying electrons and holes}
\label{chap06:2}

The conventional explanation for mobile electrons and holes in a KI at low $T$
is thermal excitation across the gap. The fact that the
zero-field resistivity of YbB$_{12}$ can be fitted using an Arrhenius formula
over the range $6 \lesssim T \lesssim 50$~K \cite{Xiang2018SdH}, albeit
with two different gaps for lower and higher temperature regimes, is
good evidence that this mechanism does, in fact, contribute.
However, below $T\approx 2~$K, the resistivity saturates, suggesting that as
the number of thermally-activated carriers declines as $T \rightarrow 0$, another mechanism
starts to dominate.

Various origins have been proposed for the neutral quasiparticles in YbB$_{12}$, including 3D Majorana fermions (a consequence of the mixed-valence of Yb)~\cite{erten2017skyrme,VarmaMajorana},
Majorana polarons~\cite{MajoranaPolaron}, spinons~\cite{Sodemann,RaoCyclotron} and related composite (fermionic) excitons~\cite{Chowdhury} {\it etc.}. The identity of the neutral fermions in YbB$_{12}$ is beyond the scope of the present work; however, we point out that all of the above approaches represent stationary states of an infinite, perfectly periodic crystal. The introduction of surfaces, grain boundaries, impurities and other defects (and higher $T$) will result in the neutral fermion states possessing finite lifetimes. As the proposed neutral quasiparticle models represent superpositions of electrons and holes, a likely decay/scattering route is into these charge-carrying in-gap states.
An immediate consequence of such a mechanism is that the {\it most perfect} crystals of YbB$_{12}$
will show the {\it highest} values of saturated, low$-T$ resistivity ({\it i.e.,} the exact opposite of the situation in conventional metals!). If one gauges crystal perfection using the
absolute amplitudes of the KI-state SdH and dHvA oscillations ({\it i.e.,} the larger the oscillations, the more perfect the crystal), this is indeed true for all of the samples we examine \cite{Xiang2018SdH,Sato2019thermal}.

It is likely that when a neutral quasiparticle scatters off lattice defects/impurities, it will decay into an electron-hole pair. Such a process functions as indirect scattering between neutral and charged excitations; a change in the neutral quasiparticle DOS will modify the scattering rate and thus be reflected in the number of charge carriers contributing to electrical transport. In addition, direct scattering between neutral and charged quasiparticles will occur, most likely via a Baber process. (Baber processes describe the interband scattering between multiple electron reservoirs with different masses \cite{Baber}). The direct scattering rate should be proportional to the joint DOS of the bands involved. Consequently, both scattering mechanisms \cite{footnote} will convert features in the field-dependent DOS of the neutral quasiparticles to signatures in the electrical transport measurements (including the Lifshitz transitions and SdH effect, as we will discuss below). A more detailed description of the scattering mechanisms in the two-fluid model must await further theoretical work.

\subsection{Temperature- and Field-dependent Hall effect: Surface vs. Bulk Conduction}
\label{chap06:3}

Whereas several measurements~\cite{Xiang2021pulsed,Xiang2018SdH,FIBYbB12Sato,Sato2019thermal}
demonstrate that the neutral fermions and the thermally-activated electrons and holes inhabit the bulk of YbB$_{12}$ (and are therefore definite intrinsic properties), studies of the focused-ion-beam (FIB)-etched YbB$_{12}$ samples reveal a substantial surface contribution at low $T$ and low $H$~\cite{FIBYbB12Sato}: as $H$ increases, charge transport in YbB$_{12}$ undergoes a crossover from surface-dominated to bulk-dominated, whilst the SdH effect arises only from bulk states.

Taking the above points into account, we assign the low-field Hall anomaly shown in Fig.\,1 to the surface conduction channel, whereas the $T$-dependent Hall signal at higher fields
is associated with the bulk.
Consequently, the analysis in Sec.\,\ref{chap03} points to the incoherent nature of the extended surface states in YbB$_{12}$: their behavior is analogous to that of the strange metal phase in correlated-electron systems~\cite{PutzkeHall,HallBaFeAsP,HallFeSeS}.
It is likely that several factors contribute to the loss of quasiparticle coherence at the surface:
(i)~there will be scattering of the surface charge carriers into the DOS
associated with the charged fermions in the
bulk of the material(Similar surface-bulk scattering processes have been widely reported in candidate topological insulators~\cite{Bi2Se3scattering,ParallelMR} and Weyl semimetals \cite{QPIWTe2}); (ii)~in order to the maintain neutral-fermion density, charged fermions at the surface will be subsumed into the bulk itinerant neutral states (the reverse of the process that creates them); and (iii)~there is also expected to be scattering due to magnetic fluctuations \cite{AFMfluc}, which can be enhanced significantly at the surface where the Kondo interaction is weakened compared to the bulk~\cite{SmB6SpinExcitation,Kondobreakdown}.

We argue that the abundance of scattering process (iii), implicit in the Hall expression Eq. 1 and resembling those in quantum critical metals, is a major ingredient in the absence of SdH oscillations in the FIB microstructures \cite{FIBYbB12Sato}. Although the conducting surface states can also experience scattering with charge-neutral quasiparticles, strong scattering from the magnetic fluctuations at the surface may be the dominant process and perhaps cause a strange metallic state at the surface. As a result, the neutral quasiparticle DOS has only a small effect on the surface conduction, so that the SdH effect is suppressed below our experimental resolution. Other factors, such as the destruction of neutral quasiparticle bands by Kondo breakdown at the surface \cite{VarmaMajorana,Kondobreakdown}, or the excess surface defects introduced by the FIB etching, may also contribute to a reduced neutral quasiparticle lifetime, thus suppressing SdH oscillations.

\subsection{Putative Lifshitz transitions}
\label{chap06:4}

As increasing magnetic field suppresses the Kondo gap \cite{SugiyamaHighField},
the bulk transport channels come into play.
The first characteristic field, manifested by the upturn in $\tau$ and a cusp in the MR~\cite{Xiang2018SdH}, occurs at $\mu_0H_1$ = 19.6\,T. Given that the ground state of Yb$^{3+}$ in YbB$_{12}$ has $\Gamma_8$ symmetry with $g_Jm_J =2.1$ (here $g_J$ is the Land\'{e} $g$-factor and $m_J$ is the magnetic quantum number for angular momentum)~\cite{Terashima2017JPSJ}, the Zeeman energy $E_z$ of the $\Gamma_8$ multiplets at $H=H_1$ is 2.39~meV$\approx \Delta_2/2$, where $\Delta_2 = 4.70~$meV is the low-temperature transport gap determined from $\rho_{xx} (T)$ \cite{Xiang2018SdH}. The close correspondence
of these energies suggests that $H_1$ may be related to the Zeeman splitting of the
crystal-electric-field (CEF) multiplets of the Yb $f$-states. However, $H_1$ is not a real gap-closing ({\it i.e.,} I-M transition)
field because there is no evidence for a sudden DOS enhancement at this field in
transport data~\cite{Xiang2018SdH} and heat capacity~\cite{Terashima2018PRL} measurements (consequently, $\Delta_2$ might be the energy separation between an in-gap impurity level and one of the band edges).
The Hall coefficient data merely show a downward kink at $H_1$,
suggestive of an inflection in DOS$(H)$~\cite{YbRh2Si2Lifshitz}.
These features are consistent with a Lifshitz transition, presumably in the FS
of the charge-neutral fermions, that is related to the Zeeman shift of the CEF levels.
Though we are not able to assign $H_2$ to possible CEF level splitting effects, the distinct SdH frequencies detected below and above this field strongly imply that it can be another putative Lifshitz transition.
At these transitions, the signatures in electrical transport may originate from abrupt changes
of the interband scattering rate (which depends on the DOS of the neutral fermions)
at the Lifshitz transitions.

Further studies of the field-dependent gap evolution in YbB$_{12}$ will provide insights into the nature of the characteristic fields $H_1$ and $H_2$; at this point, we propose that the Zeeman shift of the CEF levels causes the gap narrowing under magnetic fields, which modifies the dispersion of the in-gap neutral quasiparticle bands and/or drives a relative energy shift between these bands and the chemical potential. Successive (putative) Lifshitz transitions can happen during this process before the complete closing of the charge gap. We mention that whilst field-induced CEF level mixing can also lead to abrupt FS reconstructions \cite{CEFmixing}, such effect remains very weak in YbB$_{12}$ up to $\gtrsim$40\,T \cite{KuriharaElastic} and thus is likely to be irrelevant here (See Supplemental Material \cite{Supplement} for more information).

\subsection{Mechanism for the quantum oscillations}
\label{chap06:5}

\begin{figure}[htbp!]
\centering
\includegraphics[width=0.39\textwidth]{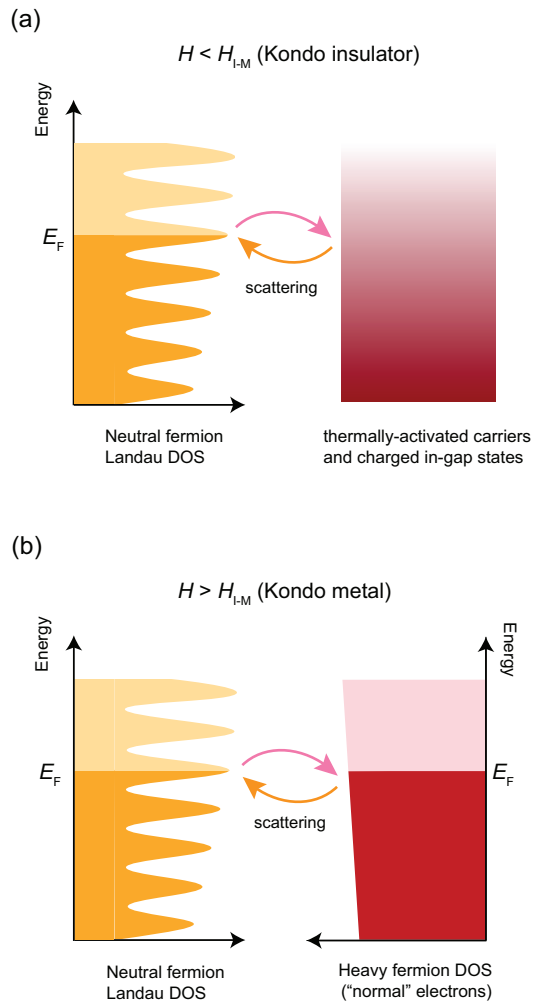}
\caption{Schematic of the two-fluid picture for transport in YbB$_{12}$;
a neutral-quasiparticle Fermi surface coexists with
more conventional charge carriers.
Close to $E_{\rm F}$, the neutral quasiparticles may likely be represented
as a linear, equal combination of canonical particle and anti-particle operators.
The linear coupling to the magnetic vector potential {\bf A} will change their
wave-vector({\bf k})-dependent energy to
$E=\frac{1}{2}[\varepsilon ({\bf k}-\frac{\rm e}{\hbar}{\bf A})+\varepsilon ({\bf k}+\frac{\rm e}{\hbar}{\bf A})]$.
Thus, any linear response to {\bf A} ({\it e.g.} electrical conductivity) will be
zero, but effects quadratic in {\bf A} ({\it e.g.,} Landau quantization, and therefore the dHvA effect) will occur.
 (a)~In the KI state, the charge carriers are primarily thermally-activated
 electrons and
holes plus (for $T\lesssim 2~$K) charged in-gap states including surface states, bulk impurity levels, etc. The rate of scattering between the charge carriers and the
neutral quasiparticles will depend on the available DOS; hence, the Landau-level DOS of the neutral quasiparticles will cause a field-dependent modulation of the electrical transport, leading to SdH oscillations.
(b)~In the KM state, fermions scatter between charge-neutral Landau levels
and heavy-fermion metal bands;
again the scattering occurs at a rate given by the joint fermion DOS.
As the neutral quasiparticle DOS oscillates with $H$, so will the transport parameters
of the heavy electrons.
(The heavy electrons are thought to have a mass $\approx 100m_0$ and so their quantum oscillations will be thermally smeared out even at the lowest experimental $T$
thus far~\cite{Xiang2021pulsed}.)
}
\vspace{-7mm}
\label{PippScat}
\end{figure}

The most important question in our study of YbB$_{12}$ is the origin of the
quantum oscillations in the Kondo-insulator state. In our two-fluid picture (Fig~\ref{PippScat}),
this is attributed to Landau
quantization of the charge-neutral FS and the interband scattering between
charged and neutral-fermion states~\cite{Xiang2021pulsed}.
In the present paper, our observation of the $T^{\alpha}$-dependence $(-2 \lesssim \alpha \lesssim -1.5)$
of high-field $\rho_{xx}$ data further supports this explanation by confirming
that the charge carriers are not in a coherent FL state and therefore cannot {\it cause}
the observed quantum oscillations. Instead, the {\it neutral-fermion Landau levels}
modulate the scattering rate (Fig.~\ref{PippScat}) via the processes introduced in Sec.\,\ref{chap06:2}, leading to the SdH
effect in $\rho_{xx}$.

We again emphasize that, although the $T$-dependence of the SdH amplitudes remain complex due to the plethora of scattering processes, the excellent fit of the dHvA amplitudes to the LK formula [Fig. 7(a), see also Refs~\cite{Xiang2018SdH,Xiang2021pulsed}] is a definitive signature of FL-like behavior. This explicitly supports the gapless neutral-fermion interpretation and indeed helps us to rule out most of the alternative scenarios. Numerous hypothetical models predict that quantum oscillations can emerge in {\it gapped} electronic systems through the field-induced tunneling effect~\cite{Knolle2015, Cooper2021}, exciton formation~\cite{Knolle2017}, periodic variation of the gap width~\cite{InvertedTI,CPasymmetry,ModifiedLK,Peters,KI_Cyclotron}
or Kondo-coupling strength~\cite{EnhancedQOs}, or Landau quantization of band edge states in conjunction with a sharp slope change in a fully-filled band~\cite{FermiSea}.
All of these {\it gapped} models cannot explain the experimental
$T$-dependence of the quantum oscillations in YbB$_{12}$ because they predict non-LK behavior of the oscillation amplitudes or even a complete absence of the SdH effect.

Another question is whether or not a disorder-induced nodal semimetal phase could contribute to or cause the quantum oscillations~\cite{nonHermitian,HarrisonNode}. In such a phase, the (incoherent) non-Hermitian in-gap states, despite being characterized by a finite quasiparticle lifetime, can form Landau levels under magnetic fields and produce quantum oscillations with LK-like-behavior~\cite{nonHermitian}. At present we cannot completely rule out this possibility, since the $T^{\alpha}$-dependence of $\rho_{xx}$ may also suggest nodal transport, as discussed in Sec.\,\ref{chap05:2}. However, this scenario is unlikely because of its prediction of an apparent mass term in the LK expression that is determined by the scattering rate~\cite{nonHermitian}. Immediately above $H_{\rm I-M}$, the boosted population of charged fermions \cite{Terashima2018PRL} should change the scattering rate significantly, yet the mass obtained from LK fits is still very close to that below $H_{\rm I-M}~(\approx 7m_0)$~\cite{Xiang2021pulsed}. This strongly suggests that the scattering rate does not affect the measured effective mass.. Hence, we believe that the incoherent charge carriers in the KI state of YbB$_{12}$ are incapable of developing Landau quantization themselves and the SdH effect comes from their scattering with the Landau-quantized neutral fermions (see Fig.~\ref{PippScat}).

We also note that quantum oscillations in the resistivity and magnetization of YbB$_{12}$ have been observed by a Cambridge group~\cite{HsuLiu,CambridgeYbB12}. Some of the orbit areas and cyclotron masses that they reported are similar to our results. The Cambridge group further mention that, unlike SmB$_{6}$ in which the quantum oscillation patterns match the large FSs in the
absence of hybridization~\cite{tan2015unconventional,Hartstein2018Fermi}, the quantum oscillations in YbB$_{12}$ stem from small FS pockets of neutral quasiparticles.
Despite these consistencies, their samples show remarkably low resistivity for a KI:
the low-$T~\rho_{xx}$ of only $\sim 0.01~\Omega$cm~\cite{HsuLiu}, two order of magnitudes smaller than that of our samples, is likely attributable to a higher concentration of impurities (see discussion above and Ref.~\cite{Li2021review}). Therefore, owing to the sensitivity of the Kondo gap and the various low-energy scattering effects to impurities~\cite{MItransition,DonorMixedValence,FragilityKondo,MagneticImpurity}, the phenomena observed in the two types of YbB$_{12}$ sample, whilst being part of the same unified picture, may differ substantially.

\section{Conclusions}

In summary, we report a detailed study of electrical transport and magnetic torque in the Kondo insulator YbB$_{12}$ under the extreme conditions of low temperatures and high magnetic fields.
A low-field Hall anomaly with an unusual nonlinear$-H$ profile is revealed for $T \lesssim 1.5~$K;
this cannot be explained by a standard two-band Drude model, and is instead
assigned to the contribution from strange-metal-like non-FL states (most likely residing on the surfaces) without full quasiparticle coherence.
At $\mu_0H_1 = 19.6~$T, a prominent signature appears in both MR and $\tau$, coinciding with a slope change in the Hall coefficient. A similar, but less prominent, signature is observed at $\simeq$30.4-31.3\,T with a weak sample dependence. We interpret these as putative field-induced Lifshitz transitions that, in a similar manner to the SdH oscillations in YbB$_{12}$, occur in the FS of exotic charge-neutral fermions and are manifested in transport via scattering with the charge carriers (Fig.~\ref{PippScat}).
We further show that the apparent different $T-$dependences of the dHvA and the SdH oscillation amplitudes is mainly caused by a $T^{\alpha}$-behavior of the background resistivity, with $\alpha$ being
$H-$dependent~($-2 \lesssim \alpha \lesssim -1.5)$.
This suggests the high-field charge transport in YbB$_{12}$ is subject to unusual scattering mechanisms that are very different from those in normal metals.

The above results strongly support a two-fluid model involving of two type of fermions: one set behaving like a conventional FL, yet not carrying charge; the other comprising charge carriers that exhibit incoherent, non-FL-like electrical transport. The interplay between the two fluids gives rise to the unusual transport phenomena including the possible Lifshitz transitions and the SdH effect (in a KI state). While our study provide solid evidence for the presence of charge-neutral quasiparticles in YbB$_{12}$, further investigation is underway to clarify their exact nature.

\addcontentsline{toc}{section}{ACKNOWLEDGMENTS}
\section*{ACKNOWLEDGMENTS}

We thank T. Wu for fruitful discussions. The work at Michigan was supported by the National Science Foundation under Award Nos. DMR-1707620 and DMR-2004288 (transport measurements), by the Department of Energy under Award No. DE-SC0020184 (magnetization measurements), by the Office of Naval Research through DURIP Award No. N00014-17-1-2357 (instrumentation). The work at Kyoto was supported by Grants-in-Aid for Scientific Research (no. JP18H05227) and by JST CREST (JPMJCR19T5). A portion of this work was performed at the National High Magnetic Field Laboratory (NHMFL), which is supported by National Science Foundation Cooperative Agreement No. DMR-1644779 and the Department of Energy (DOE). JS acknowledges support from the DOE BES program ``Science at 100~T'', which permitted the design and construction of much of the specialised equipment used in the high-field studies. The experiment in NHMFL is funded in part by a QuantEmX grant from ICAM and the Gordon and Betty Moore Foundation through Grant GBMF5305 to Z.X., L.C., T.A., C.T. and L.L.

\appendix
\section*{Appendix}
\label{chap07}
\addcontentsline{toc}{section}{Appendix}
\renewcommand{\thesubsection}{\Alph{subsection}}

\subsection{Low-temperature resistivity}
\label{chap07:1}

\begin{figure}[htbp!]
\centering
\includegraphics[width=0.48\textwidth]{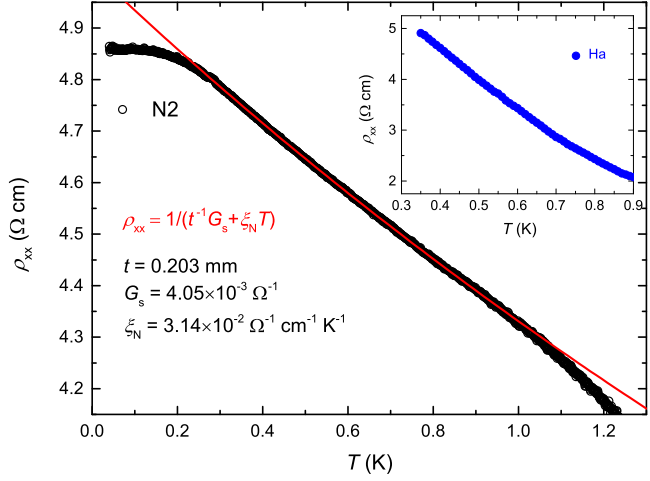}
\caption{Resistivity $\rho_{xx}$ of sample N2 as a function of temperature below $T = 1.3~$K, measured in a dilution refrigerator. Between 0.3~K and 1.0~K, $\rho_{xx}$ of sample N2 can be fitted by
$\rho_{xx}$ = ($G_{\rm s}$/$t$+$\xi_{\rm N} T$)$^{-1}$ (see text) (solid line). The fitting parameters are also shown. The inset shows the low temperature resistivity of YbB$_{12}$ sample Ha.
}
\label{FigRTfit}
\end{figure}

The longitudinal resistivity $\rho_{xx}$ in sample N2 is displayed in Fig.~\ref{FigRTfit} for $T < 1.3~$K. $\rho_{xx}$ increases slowly with cooling down, appears to saturate for $T\lesssim 0.2~$K.
Note that the $T-$range shown in Fig.~\ref{FigRTfit} is within the
``resistive plateau"~\cite{Xiang2018SdH} in YbB$_{12}$ which occurs below $T\approx 2.2~$K.
The $T-$dependence of $\rho_{xx}$ between 0.25 and 1\,K can be fitted by a simple function that invokes both surface and bulk transport channels:
\begin{equation}
\rho_{xx}(T) = \frac{1}{G_{\rm s}/t + \sigma_b} = \frac{1}{G_{\rm s}/t + \xi_N T},
\label{RT}
\end{equation}
where $G_{\rm s}$ is the surface conductance, $t$ is the sample thickness and {$\sigma_{\rm b}= \xi_{\rm N} T$ is the bulk conductivity which is linear in $T$.

The fit (red line in Fig.~\ref{FigRTfit}) yields $G_s = 4.05 \times 10^{-3}~\Omega^{-1}$ (corresponding to a surface resistance of $246.9~\Omega$) and $\xi_{\rm N} = 0.0314~\Omega^{-1}{\rm cm^{-1}K}^{-1}$.
We note that Eq.\,\ref{RT} decomposes the overall resistivity into two contributions: one from a $T-$independent surface conduction, the other due to a bulk conduction which varies linearly with $T$. The same formula has been proposed to interpret the low$-T$ resistive plateau in SmB$_6$~\cite{HarrisonNode}, where the $T$-linear bulk conductivity is taken as the evidence for the formation of gap nodes due to the impurity-scattering-induced level broadening. Following the analysis in Ref.\,\cite{HarrisonNode}, the node conductivity is given as:
\begin{equation}
\xi_N = \frac{ln 2}{4\pi^3} \frac{k_B e^2 S_k}{\hbar k_0} \frac{\Gamma_0}{V^2}.
\label{node}
\end{equation}
Here, $S_k$ is the Fermi-surface area for the unhybridized conduction band, $k_0$ represents the position of the impurity-induced nodes in momentum space, whereas $\Gamma_0$ and $V$ are the energy level broadening and the hybridization potential, respectively. By using the electrical-transport results ($\Delta_2 = 3.2-4.7~$meV), quantum oscillation data ($F \approx 700~$T, $m^*/m_0 \approx 7$)~\cite{Xiang2018SdH,Xiang2021pulsed} and ARPES results~\cite{ARPESYbB12} in YbB$_{12}$,
we estimate $k_0 \approx 0.2~{\rm {\AA}}^{-1}$ and $V \simeq 9.2-11.5~$meV. Therefore, the fitted
$\xi_{\rm N} = 0.0314~\Omega^{-1}{\rm cm^{-1} K}^{-1}$ corresponds to
$\Gamma_0 S_k \approx 7-11 \times 10^{-4}~{\rm J m}^{-2}$.
To close the hybridization gap at $k_0$ by node formation, the level broadening needs
to be $\Gamma_0 \geq 2V \approx 20~$meV.
If the system is actually driven into a nodel semimetal phase by impurity scattering,
the value of $S_k$ would be only $0.2-0.3~{\rm nm}^{-2}$ (corresponding Fermi vector
$k_{\rm F} \lesssim 0.3\,{\rm nm}^{-1}$).
Such a small Fermi surface area obviously contradicts that observed in our quantum oscillation
measurements ($k_{\rm F} \simeq 1.4-1.7~{\rm nm}^{-1}$)~\cite{Xiang2018SdH}
and is even more inconsistent with the calculated large $d$-electron pockets for
unhybridized YbB$_{12}$~\cite{HsuLiu}.
We thus conclude that the behavior of $\rho_{xx}(T)$ in YbB$_{12}$ cannot be
explained by the impurity-induced nodal in-gap states.

\begin{figure}[htbp!]
\centering
\includegraphics[width=0.48\textwidth]{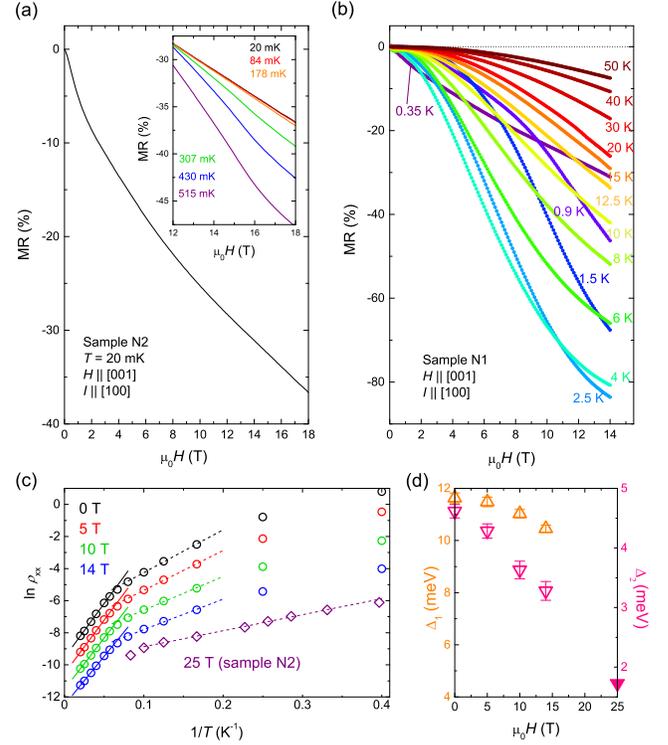}
\caption{(a) Transverse MR ($\rho_{xx}$($H$)-$\rho_{xx}$($0$))/$\rho_{xx}$(0)$\times$100$\%$ measured in sample N2 at $T$ = 20 mK, with magnetic field up to 18 T applied along [001] and current $I \parallel$ [100]. Inset exhibits transverse MR of sample N2 between 12 and 18 T at various temperatures. (b) Transverse MR of sample N1 up to 14 T for temperatures between 0.35 and 50 K. (c) The Arrhenius plot $\ln\rho_{xx}$ versus 1/$T$ for sample N1 under different magnetic fields up to 14\,T (hollow circles). The slopes of the linear fits for temperature ranges 17.5\,K $\leq T \leq$ 50\,K (solid lines) and 6\,K $\leq T \leq$ 12.5\,K (dashed lines) gives the value of gap widths $\Delta_1$ (up triangles) and $\Delta_2$ (down triangles) in (d), respectively. Data taken in sample N2 at 25\,T is also displayed with the linear fit between 3 and 8\,K yields $\Delta_2$ = 1.725 meV (solid down triangle in (d)). The plots are shifted vertically by 1 for clarity.
}
\label{FigMR}
\end{figure}

Moreover, our recent transport study in FIB-etched YbB$_{12}$ microstructures also confirms the validity of Eq.\,\ref{RT} in describing the low-$T$ resistivity. However, in that case, the term $\xi_N$ was revealed to be thickness-dependent~\cite{FIBYbB12Sato}. This provides further evidence that the $T$-linear term is not solely associated with the bulk states. We also mention that the low-$T$ $\rho_{xx}$ shows different behavior between batches. As displayed in the inset of Fig.\,8, $\rho_{xx}$ in sample Ha exhibits a stronger $T$-dependence compared to sample N2 below 1~K. $\rho_{xx}(T)$ of sample Ha cannot be successfully fitted by Eq.\,\ref{RT}. These differences presumably reflect the different surface conditions on the various samples, since the low-temperature, low-field electrical transport in YbB$_{12}$ seems to be dominated by surface contributions~\cite{FIBYbB12Sato}.

We mention that our experimental results for YbB$_{12}$ microstructures \cite{FIBYbB12Sato} cannot completely rule out the existence of bulk conducting channels at zero temperature. A recent theoretical work predicts that a low-$T$ resistivity saturation can originate from bulk states in a ``quantum-conduction" regime beyond the conventional Boltzmann picture \cite{Pickem}. Such scenario, however, is to be tested for YbB$_{12}$ only after the dominating surface conduction at low $T$ (as confirmed by the highly thickness-dependent resistivity below the saturation \cite{FIBYbB12Sato}) can be isolated using unique measurement configurations, {\it e.g.}, the double-sided Corbino device \cite{DoubleCorbino}. We emphasize that the presence of a small residual bulk conduction will not invalidate the main conclusions of this work.

\subsection{Magnetoresistance}
\label{chap07:2}

\begin{figure}[htbp!]
\centering
\includegraphics[width=0.45\textwidth]{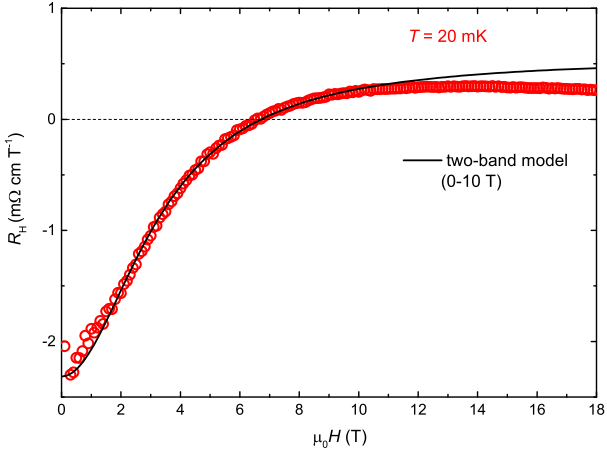}
\caption{Fitting of the field dependence of the Hall coefficient $R_{\rm H}(H)$ below 10 T using a two-band model with the constraint from the zero-field resistivity (Eq.\,\ref{KimHall} in Appendix Sec.\,\ref{chap07:3}). $R_{\rm H}(H)$ is measured in sample N2 at $T$ = 20 mK.
}
\label{FigHallSCM1}
\end{figure}

Charge transport in the ground state of YbB$_{12}$ is characterized by a negative MR, which reaches $-37\%$
at $\mu_0H = 18~$T ([Fig.\,10(a)]. This negative MR was previously believed
to stem from the field-suppression of the Kondo gap through the Zeeman splitting of the
$f$-electron multiplets~\cite{SugiyamaHighField,Terashima2017JPSJ}.
Recently we demonstrated that for $T <1~$K, the MR in the low$-H$ region (up to 8\,T)
can be accounted for by the Hikami-Larkin-Nagaoka model for weak antilocalization,
while the gap suppression explanation works for the negative MR at higher fields~\cite{FIBYbB12Sato}. Nevertheless, the negative MR corroborates that the electrical-transport properties of
YbB$_{12}$ cannot be described by the classical two-band model that predicts an
enhanced positive orbital contribution to the MR \cite{AliWTe2}.

As shown in the inset of Fig.\,10(a) and Fig.\,10(b), the negative transverse MR (measured with $H \perp I$) in YbB$_{12}$ first increases upon warming, reaches a maximum of $-83.5\%$ at 14\,T and $T = 2.5~$K, then gradually decreases as $T$ increases further.
No positive MR is observed up to 50\,K.
We conclude that the low-$T$ charge transport in YbB$_{12}$ exhibits distinct nonmetallic behavior and thus can not be explained within the frame of 3D band-like quasiparticle transport theory. Fig.\,10(c) displays Arrhenius plots $\ln\rho_{xx}$ versus $1/T$ for different magnetic fields. As noted previously~\cite{Xiang2018SdH},
two gap widths ($\Delta_1$ and $\Delta_2$) can be obtained from the linear fits within
the temperature ranges $17.5~{\rm K} \leq T \leq 50~$K and $6~{\rm K} \leq T \leq 12.5~$K, respectively.
The field dependence of the two gap widths is shown in Fig.\,10(d). It appears that the low-temperature transport gap $\Delta_2$ is more severely affected by magnetic field, decreasing by roughly one third at 14\,T
compared to the $\approx 10\%$ suppression for $\Delta_1$. This phenomenon is consistent with earlier reports in both YbB$_{12}$~\cite{SugiyamaHighField} and SmB$_6$~\cite{Shahrokhvand}.

\begin{figure*}[htbp!]
\centering
\includegraphics[width=0.75\textwidth]{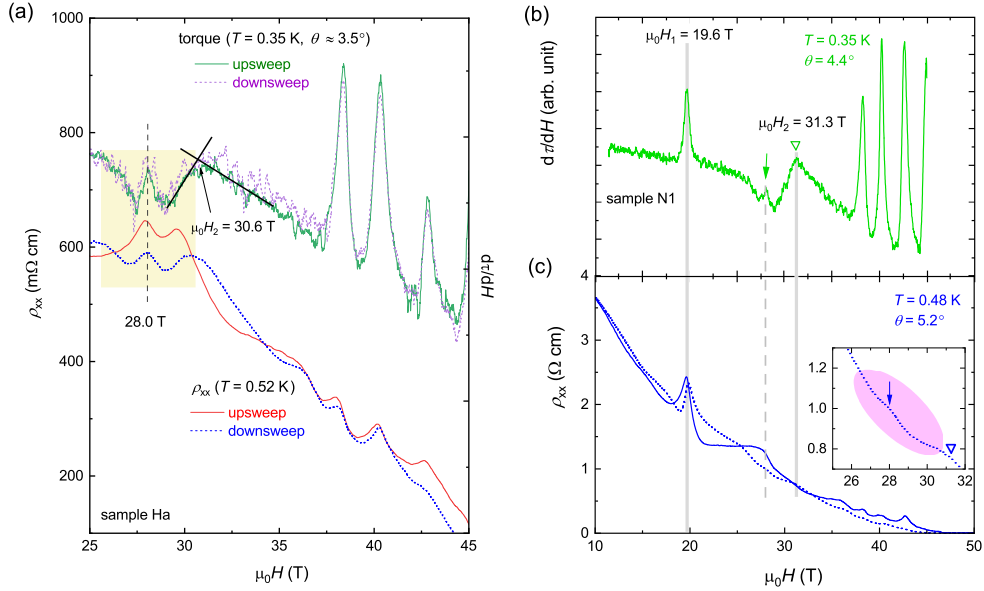}
\caption{(a)Magnetic field dependence of MR and d$\tau$/d$H$ for YbB$_{12}$ sample Ha, with the magnetic field close to the [001] direction. Solid and short-dashed lines represent data curves taken during field upsweep and downsweep, respectively. To be noted, two wiggles appear between 25.5\,T and 30.5\,T (yellow shaded region) with the peaks and valleys well aligned between $\rho_{xx}$ and d$\tau$/d$H$. These are quantum oscillations. In particular, the signature at 28\,T (vertical dashed line) should be an SdH peak. An abrupt slope change of d$\tau$/d$H$ at ~30.6 T, on the other hand, is likely to be a Lifshitz transition. Magnetic field dependence of (b) d$\tau$/d$H$ measured at $T$ = 0.35\,K for tilt angle $\theta$ = 4.4$^\circ$, (c) MR at $T$ = 0.48\,K for $\theta$ = 5.2$^\circ$. Data in (b) and (c) were measured in sample N1. Solid and short-dashed lines in (c) are upsweeps and downsweeps, respectively. Characteristic fields $\mu_0H_1$ = 19.6\,T (manifested by prominent peaks in both d$\tau$/d$H$ and MR) and $\mu_0H_2$ = 31.3\,T (a sharp kink in d$\tau$/d$H$) are denoted by vertical solid gray lines. The feature appearing at 28\,T in both d$\tau$/d$H$ and MR (arrow and dashed vertical line) originates from quantum oscillations. Inset of (c) is an expanded view of $\rho_{xx}$($H$) between 25 and 32\,T. The light magenta shaded area indicates the field range at which hints of weak SdH effect can be seen on the downsweep. The down triangle marks the position of $\mu_0H_2$.
}
\label{FigSampleN1}
\end{figure*}

At 25\,T, the linear fit does not work in the same temperature window for $\Delta_2$, yet the slope between 3\,K and 8\,K gives a gap width of 1.725~meV, {\it i.e.,} less than 40$\%$ of the zero-field value [Fig\,10(c)]. We add this point to Fig.\,10(d) as a solid triangle. With $H$ further increasing, however, the linear section in $\ln\rho_{xx}$ versus $T^{-1}$ cannot be determined and $\rho_{xx}$ develops the $T^{\alpha}$ ($\alpha <$ 0) behavior seen above 35\,T (Fig.\,6).

\subsection{Two-band analysis of the Hall coefficient}
\label{chap07:3}

In most cases, a nonlinear $\rho_{xy}(H)$ is taken as evidence for a multiband carrier system comprising electrons and holes. Here, we define a field-dependent Hall coefficient $R_{\rm H} \equiv \rho_{xy}/\mu_0H$ and plot it as a function of $H$ for $T=20~$mK in Fig.\,1(b). For $H\rightarrow 0$, $R_{\rm H}$ is negative; as $H$ ramps up, $R_{\rm H}$ is quickly suppressed, crossing zero at $\approx 6.6~$T. It then shows a very weak downward bending for $\mu_0H >10~$T. We first focus on the low$-H$ region below 10\,T where the field dependence of $R_{\rm H}$ is most evident. Standard Boltzmann-transport theory provides an expression for $R_{\rm H}$ in a two-band system \cite{AshcroftMermin,Hurd_Hall}:
\begin{equation}
R_{\rm H}(H) = \frac{1}{e}\frac{(n_h\mu_h^2+n_e\mu_e^2)+(n_h-n_e)\mu_h^2\mu_e^2(\mu_0H)^2}{(n_h\mu_h+n_e\mu_e)^2+(n_h-n_e)^2\mu_h^2\mu_e^2(\mu_0H)^2},
\label{twobandHall}
\end{equation}
where $n_h$ is the density of holes, $n_e$ the density of electrons, $\mu_h$ the hole mobility and $\mu_e$ the electron mobility. We apply Eq.\,\ref{twobandHall} to the $R_{\rm H}(H)$ values shown in Fig.\,1(b) with the parameters $n_h$, $n_e$, $\mu_h$ and $\mu_e$ further constrained using the value of the zero-temperature resistivity $\rho_{xx}(0) = (e n_h\mu_h) + e n_e\mu_e)^{-1} = 4.85~\Omega$cm (Fig.\,9). Following the discussion in Ref.\,\cite{Kim_Hall}, we can write $R_{\rm H}$ as:
\begin{equation}
R_{\rm H} = \rho_{xx}(0)\frac{a+b(\mu_0H)^2}{1+c(\mu_0H)^2},
\label{KimHall}
\end{equation}
where $a = f_h\mu_h-f_e\mu_e$, $b = -(f_e\mu_h-f_h\mu_e)\mu_h\mu_e$, $c = (f_h\mu_e-f_e\mu_h)^2$ and $f_{h,e} = (n_{h,e}\mu_{h,e})/(n_h\mu_h+n_e\mu_e)$. Hence the number of fitting parameters is reduced to three: $\mu_h$, $\mu_e$ and $f_h$ ($f_e$ = 1-$f_h$). The fitting curve using Eq.\,\ref{KimHall} is displayed in Fig.\,11. From this fit, we obtain the following parameters:
$n_h$ = (1.13$\pm$0.01)$\times$10$^{18}$ cm$^{-3}$, $n_e$ = 8.20$\pm$0.05)$\times$10$^{11}$ cm$^{-3}$,
$\mu_h$ = 1.1$\pm$0.1 cm$^2$V$^{-1}$s$^{-1}$ and $\mu_e$ = 3045$\pm$32 cm$^2$V$^{-1}$s$^{-1}$. Notably, the fitting parameters point toward the existence of a very small electron pocket with relatively high mobility. Taking into account possible surface conduction, we can assume that the electron pocket, which has a lower carrier density, is associated with 2D surface states, whilst the holes inhabit the bulk of the sample. However, the fitted $n_e$ converts to a 2D carrier density $n_e^{2D}$ = (1.66$\pm$0.02)$\times$10$^{10}$ cm$^{-2}$ which is three orders of magnitude lower than  $n_e^{\rm 2D} = 5.16\times 10^{13}~{\rm cm}^{-2}$ determined from the ARPES results~\cite{ARPESYbB12} ($k_{\rm F} \simeq 0.18~{\rm \AA}^{-1}$ for the surface state on the (001) surface; the possible lifting of spin degeneracy is not considered here). Indeed, the small $n_e^{\rm 2D}$ yielded by the two-band fit corresponds to $k_{\rm F} = 0.0323~$nm$^{-1}$ and a quantum oscillation frequency $F = 0.343~$T under the assumption of an isotropic FS. While we can not completely exclude the existence of such tiny surface FS pockets, no further supporting evidence for them has been revealed by ARPES~\cite{ARPESYbB12} and transport~\cite{Xiang2018SdH,FIBYbB12Sato}. Additionally, the two-band model (Eq.\,\ref{KimHall}) does not give a perfect fit for the evolution of $R_{\rm H}$ from 2\,T to $\approx 5~$T (i.e., around the maximum of
$\rho_{xy}(H)$). It also fails to track the behavior of $R_{\rm H}$ above 10\,T.

We further mention that it is impossible to perform simultaneous fits to
$\rho_{xy}(H)$ and $\rho_{xx}(H)$~\cite{HallFeSeS,RourkeYBCO} in YbB$_{12}$:
whereas the two-band Drude model predicts a positive MR due to the orbital motion of carriers
in magnetic field, in YbB$_{12}$ the MR is always negative (Appendix Sec.\,\ref{chap07:2}).
The absence of positive MR indeed strongly challenges the validity of the two-band Drude description.
The nonmetallic $\rho_{xx}(T)$ (Appendix Sec.\,\ref{chap07:1}) also clearly deviates from Drude-like behavior. Moreover, aside from the single surface FS pocket detected by ARPES~\cite{ARPESYbB12}, there is no experimental evidence for the existence of FSs of charge carriers in YbB$_{12}$. Therefore, despite the two-band models failing to yield satisfactory fits, there is no reason to add a third band to improve the fit quality. Taking all of these factors into account, the low-field Hall anomaly in YbB$_{12}$, characterized by an usual $H$-nonlinear behavior, cannot be attributed to multiband effects.

\subsection{Identification of the characteristic fields}
\label{chap07:4}

Multiple field-induced signatures appear in different physical quantities of YbB$_{12}$ below $H_{\textrm{I-M}}$ (Fig.\,4). Identification of these signatures should be assigned with caution, because quantum oscillations occurring in different field ranges (Fig. 5) can manifest as features resembling field-induced transitions. In Fig.\,12(a) we plot MR and d$\tau$/d$H$ of sample Ha together to provide insights into the nature of these signatures. Most remarkably, the features (two valleys and one peak) appearing between 25.5\,T and 30.5\,T [yellow shaded region in Fig.\,12(a)] are well aligned between the two quantities, d$\tau$/d$H$ and $\rho_{xx}$, exactly as what is expected for quantum oscillations \cite{shoenberg2009magnetic}. Hence, we identify the signature at 28\,T (arrows in Fig.\,4 and Fig\,12) to be an quantum oscillation extremum, {\it i.e.,} the SdH peak shown in Fig.\,5(b). The consistency of the location of this feature between the up-sweep and down-sweep MR curves [vertical dashed line in Fig.\,12(a)] further supports this assignment.

We identify the kink [crossing of the two black straight lines in Fig.\,12(a)] in d$\tau$/d$H$ as another putative Lifshitz transition $H_2$ in addition to $H_1$. The slightly different locations of this signature [$\simeq$30.4\,T in Fig.\,4(b) and 30.6\,T in Fig.\,12(a)] are likely due to a small change in field orientations. This transition is probably the one leading to the SdH frequency change as shown in Fig.\,5. We note that the position of $H_2$ may have a displacement for the MR upsweeps and downsweeps [Fig.\,4(c)], yet such displacement is missing for magnetic torque data [Fig.\,12(a)]. Since torque and MR were measured in static and pulsed magnetic fields, respectively, the displacement may be related to mechanisms with slow kinetics that have relaxation rates similar to the rate of change of the pulsed fields. Indeed, mismatches between the upsweeps and downsweeps in the MR measurement are more pronounced in the vicinity of the characteristic fields $H_1$ and $H_2$; this might suggest a first-order-like nature of the Lifshitz transitions that has been proposed for systems with substantial electron correlations~\cite{UPt2Si2}.

In Fig.\,12(b) and (c) we display the field dependence of d$\tau$/d$H$ and $\rho_{xx}$ of sample N1, respectively. Two sharp anomalies in d$\tau$/d$H$ are identified as the characteristic fields $H_1$ (gray vertical line) and $H_2$ (down triangle), whereas an additional signature at 28\,T represents a quantum oscillation maximum in both $\rho_{xx}$ and d$\tau$/d$H$ (arrows and the vertical dashed line). There are several points to be mentioned. The positions of $H_1$ is consistent with that in sample Ha, {\it i.e.,} this signature shows no sample dependence. On the other hand, the characteristic field $H_2$ associated with the sharp kink in d$\tau$/d$H$ (Fig.\,12(b)) is much less evident in sample Ha [Fig.\,4(b) and Fig.\,12(a)]; its position varies slightly between samples ($\mu_0H_2$ = 30.4$\pm$0.2\,T and 31.3\,T for sample Ha and N1, respecticely). The SdH oscillations between 25.5 and 30\,T [the shaded area in the inset of Fig.\,12(b)] are much weaker than those in sample Ha [Fig.\,4(c)]; this is why they were overlooked in our previous work \cite{Xiang2018SdH}.


\clearpage

\renewcommand{\thefigure}{\textbf{\arabic{figure}}}
\renewcommand{\figurename}{\textbf{Supplementary Figure\,$\!\!$}}
\renewcommand{\thetable}{\textbf{\arabic{table}}}
\renewcommand{\tablename}{\textbf{Supplementary Table\,$\!\!$}}
\renewcommand{\theequation}{S\arabic{equation}}
\newcommand{\beq}{\begin{equation}}
	\newcommand{\eeq}{\end{equation}}
\newcommand{\ud}{\mathrm{d}}

\begin{center}
{\large Supplemental Material for} \\
{\large \bf Hall anomaly, Quantum Oscillations and Possible Lifshitz Transitions in Kondo Insulator YbB$_{12}$: Evidence for Unconventional Charge Transport}

\bigskip

\author{Ziji Xiang$^{1,2}$, Kuan-Wen Chen$^1$, Lu Chen$^{1,\dag}$, Tomoya Asaba$^{1,3}$, Yuki Sato$^3$, Nan Zhang$^2$, Dechen Zhang$^1$, Yuichi Kasahara$^3$, Fumitoshi Iga$^4$, William A. Coniglio$^5$, Yuji Matsuda$^3$, John Singleton$^{6}$, Lu Li$^{1}$}

\medskip
{\it \small
\setlength{\baselineskip}{16pt}

$^1$Department of Physics, University of Michigan, Ann Arbor, Michigan 48109, USA\\
$^2$CAS Key Laboratory of Strongly-coupled Quantum Matter Physics, Department of Physics, University of Science and Technology of China, Hefei, Anhui 230026, China\\
$^3$Department of Physics, Kyoto University, Kyoto 606-8502, Japan\\
$^4$Institute of Quantum Beam Science, Graduate School of Science and Engineering, Ibaraki University, Mito 310-8512, Japan\\
$^5$National High Magnetic Field Laboratory, 1800 East Paul Dirac Drive, Tallahassee, Florida 32310-3706, USA\\
$^6$National High Magnetic Field Laboratory, MS E536, Los Alamos National Laboratory, Los Alamos, New Mexico 87545, USA
}

\end{center}
\bigskip



\newpage



\begin{figure*}[htbp!]
	\begin{center}
		\includegraphics[width=0.99\linewidth]{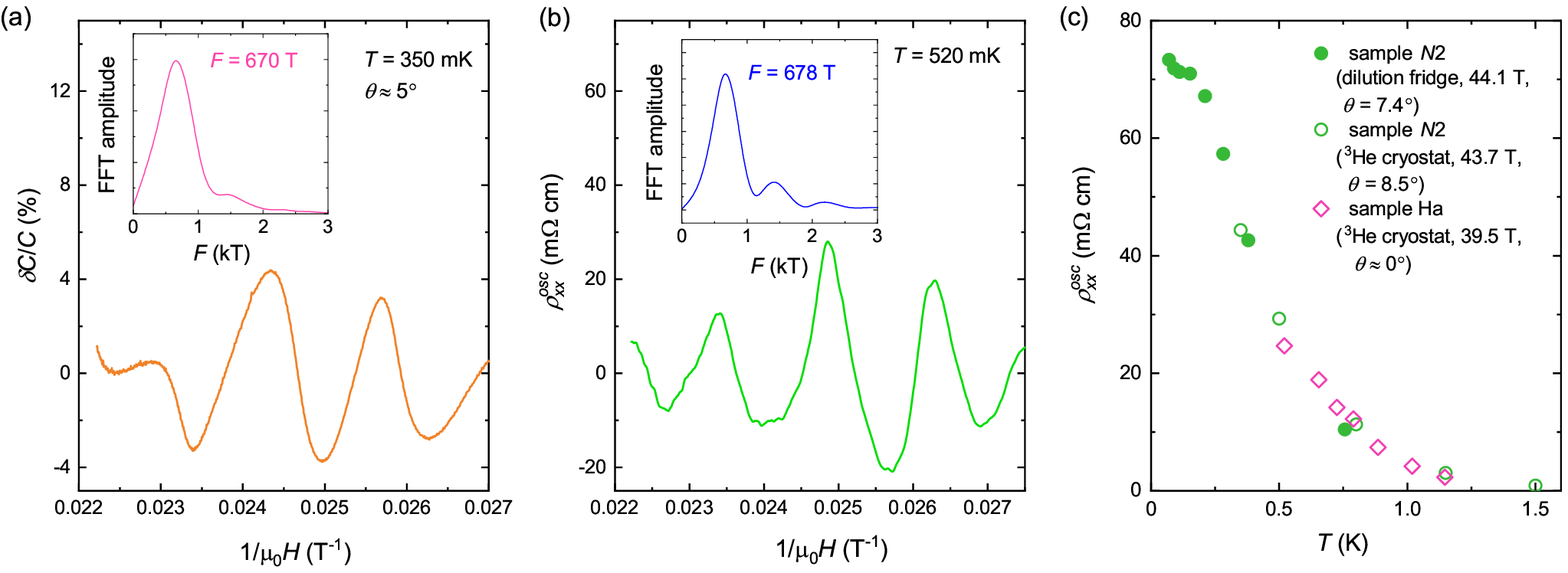}
	\end{center}
	\vspace*{-5mm}
	\caption{The quantum oscillations detected in YbB$_{12}$ sample Ha for $\mu_0H >$ 35\,T by (a) torque magnetometry (the de Haas-van Alphen effect),
           (b) magnetoresistance (MR) measurement (the Shubnikov-de Haas effect). Only the oscillatory components are shown, which are obtained by subtracting a polynomial background from the raw data. For both panels, the insets display the fast Fourier transform (FFT) spectra of the oscillations. A single frequency $F \simeq$ 670-680\,T and its higher harmonics dominate the FFT spectra without indications of additional peaks. (c) The oscillatory resistivity $\rho_{xx}^{osc}$ plotted against temperature. Here, $\rho_{xx}^{osc}$ represents the amplitude of oscillations in the raw data of MR without normalization.  Solid and hollow green circles are data measured in sample $N$2 in two different cryostats \cite{Xiang2018SdH}. Magenta diamonds are data measured in sample Ha under pulsed magnetic fields. Data normalized using the background resistivity are shown in main text Figure 7, wherein the nonmonotonic behavior below $\sim$0.5\,K is likely to be an artefact due to the normalization. It implies that the scattering between neutral quasiparticles and the charge carriers tends to saturate at the lowest temperatures, while the background resistivity continues to increases slowly upon cooling due to other mechanisms.
		}
\end{figure*}

\begin{figure*}[htbp!]
	\begin{center}
		\vspace{12pt}
		\includegraphics[width=0.8\linewidth]{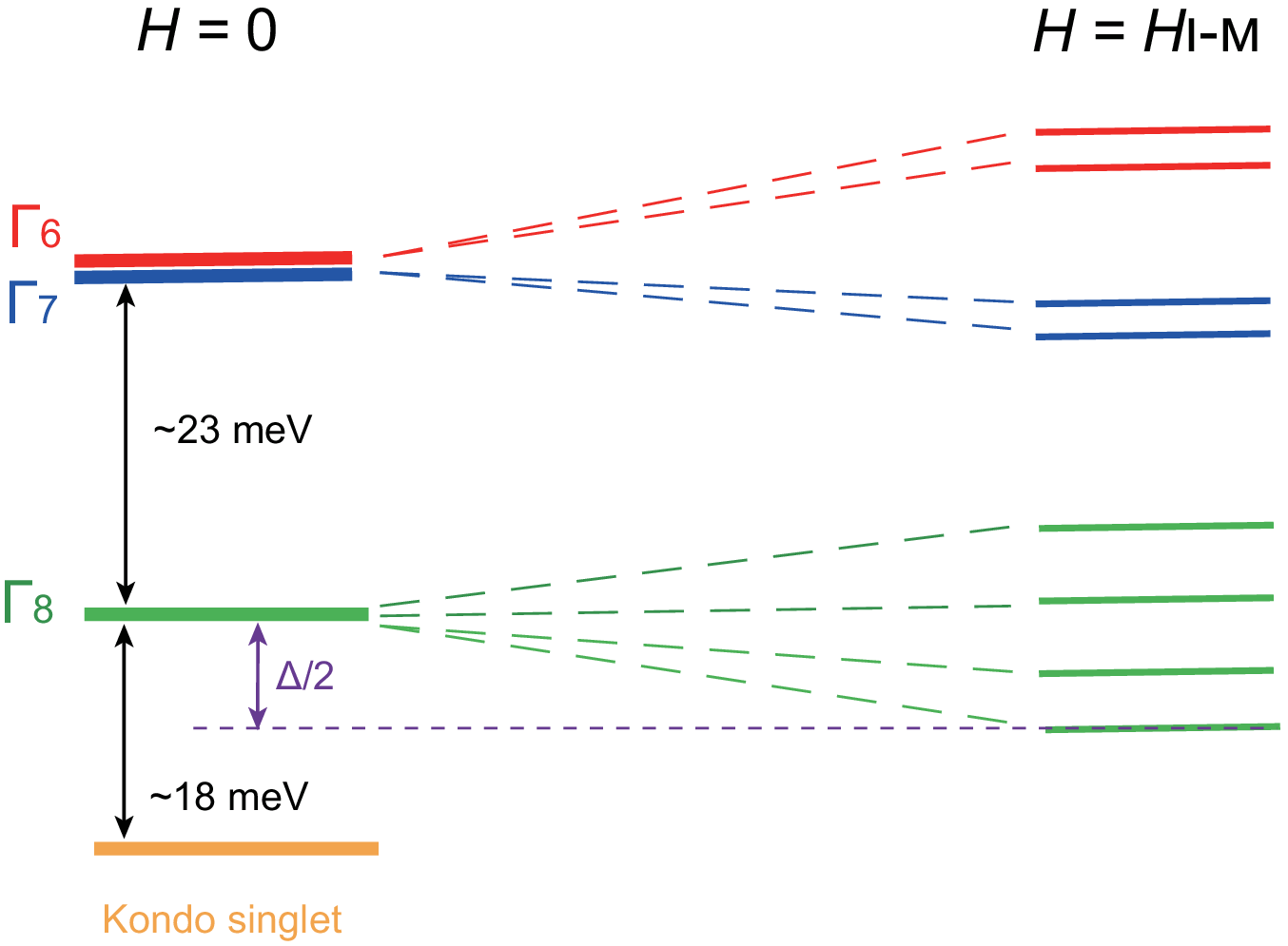}
	\end{center}
	\vspace*{-5mm}
	\caption{A schematic diagram of the low-lying crystal-electric-field (CEF) levels in YbB$_{12}$ at zero magnetic field and their field-induced splitting.
             At the gap-closing field $H_{\textrm{I-M}}$, the energy shift of the lowest CEF level of Yb$^{3+}$, {\it i.e.}, the $\Gamma_8$ multiplet, matches $\Delta/2$ where $\Delta$ is the charge gap ($\simeq$ 15 meV) \cite{Terashima2017JPSJ}. Note that the charge gap and the spin gap are not identical in YbB$_{12}$, see the Supplementary Note.
	}
\end{figure*}

\begin{center}
{\large \bf Supplementary Note: the CEF levels}
\end{center}

In the cubic lattice of YbB$_{12}$, the $J$ = 7/2 Yb$^{3+}$ (4$f^{13}$ configuration) multiplet is split into a quartet ($\Gamma_8$) and two doublets ($\Gamma_6$ and $\Gamma_7$) by the crystal electric field (CEF). The corresponding CEF level scheme is shown in Supplementary Figure 2: above the ground state, which is the Kondo singlet, the lowest CEF level of localized Yb$^{3+}$ is the $\Gamma_8$ quartet; the energy separation between the singlet ground state and the $\Gamma_8$ level ({\it i.e.,} the spin gap) is 18-20 meV \cite{AFMfluc,Nemkovski}. This is slightly larger than the charge gap $\Delta \simeq$ 15 meV \cite{Terashima2017JPSJ}. The nearly-degenerate doublets $\Gamma_6$ and $\Gamma_7$ are located at $\simeq$23 meV above $\Gamma_8$ at zero field \cite{Nemkovski}. Under magnetic fields, the Zeeman splitting of the CEF multiplets leads to energy shifts of the CEF levels. It has been pointed out that the field-induced insulator-metal transition in YbB$_{12}$ can be interpreted by the gap closing effect due to the Zeeman shift of the $\Gamma_8$ states \cite{Terashima2017JPSJ}: as illustrated in Supplementary Figure 2, the Zeeman shift of the lowest sublevel of $\Gamma_8$ reaches $\Delta$/2 at $H$ = $H_{\textrm{I-M}}$. Such scenario predicts a value of $H_{\textrm{I-M}}$ that is in agreement with the experimental results \cite{Terashima2017JPSJ}.

It is elusive at this stage whether the field-induced energy shift of the CEF levels can contribute to the observations in the present work for $H < H_{\textrm{I-M}}$, including the putative Lifshitz transitions and quantum oscillations. A recent work reports the absence of significant CEF level mixing in YbB$_{12}$ for magnetic fields up to (at least) $\sim$40\,T \cite{KuriharaElastic}. Further investigations are required to nail down the role of CEF scheme in determining the field-dependent physical properties of YbB$_{12}$.

\end{document}